\documentclass[usenatbib]{mn2e}
\usepackage{multirow}
\usepackage{rotating}
\usepackage{colortbl}
\usepackage{color}
\usepackage[fleqn]{amsmath}
\usepackage{amssymb}
\usepackage{amsfonts}
\usepackage{verbatim}
\usepackage{scalefnt}
\usepackage[percent]{overpic}

\newcommand{\kms}{\,{\rm km\,s^{-1}}}
\newcommand{\msun}{\,{\rm M_\odot}}

\newcommand{\beq}{\begin{equation}}
\newcommand{\eeq}{\end{equation}}
\newcommand{\ba}{\begin{eqnarray}}
\newcommand{\ea}{\end{eqnarray}}
\def\spose#1{\hbox to 0pt{#1\hss}}
\newcommand{\lta}{\mathrel{\spose{\lower 3pt\hbox{$\mathchar'218$}}
      \raise 2.0pt\hbox{$\mathchar"13C$}}}
\newcommand{\gta}{\mathrel{\spose{\lower 3pt\hbox{$\mathchar"218$}}
      \raise 2.0pt\hbox{$\mathchar"13E$}}}

\definecolor{grey}{rgb}{0.75,0.75,0.75}
\definecolor{Orange}{rgb}{1.0,0.5,0.15}
\definecolor{brown}{rgb}{0.7,0.25,0.0}
\definecolor{pink}{rgb}{1.0,0.5,0.5}
\definecolor{darkerred}{rgb}{0.8,0,0}
\definecolor{darkerblue}{rgb}{0,0,0.8}
\definecolor{Blue}{rgb}{0,0.08,0.65}
\definecolor{Red}{rgb}{0.65,0.08,0.05}
\definecolor{Green}{rgb}{0.15,0.45,0.25}

\newcommand{\comments}[1]{} 

\title[BHs in Horizon-AGN]{The cosmic evolution of massive black holes in the Horizon-AGN simulation}

\author[Volonteri et al.]{Volonteri, M.$^{1}$\thanks{E-mail: martav@iap.fr}, Dubois, Y.$^{1}$, Pichon, C.$^{1}$, Devriendt, J.$^{2,3}$ \\
$^1$ Institut d'Astrophysique de Paris, Sorbonne Universit\'{e}s, 
UPMC  Univ Paris 06 et CNRS, UMR 7095, 98 bis bd Arago, 75014 Paris, France\\
$^{2}$ Sub-department of Astrophysics, University of Oxford, Keble Road, Oxford OX1 3RH\\
$^{3}$ Observatoire de Lyon, UMR 5574, 9 avenue Charles Andr\'e, Saint Genis Laval 69561, France\\
}
\begin{document}

\maketitle

\begin{abstract}
We analyse the demographics of black holes (BHs) in the large-volume cosmological hydrodynamical simulation Horizon-AGN. This simulation statistically models how much gas is accreted onto BHs, traces the energy deposited into their environment and, consequently, the back-reaction of the ambient medium on BH growth.
The synthetic BHs reproduce a variety of observational constraints such as the redshift evolution of the BH mass density and the mass function.  Strong self-regulation via AGN feedback, weak supernova feedback, and unresolved internal processes result in a tight BH-galaxy mass correlation.  Starting at $z\sim2$, tidal stripping creates a small population of BHs over-massive with respect to the halo. The fraction of galaxies hosting a central BH or an AGN increases with stellar mass. The AGN fraction agrees better with multi-wavelength studies, than single-wavelength ones, unless obscuration is taken into account. The most massive halos present BH multiplicity, with additional BHs gained by ongoing or past mergers. In some cases, both a central and an off-centre AGN shine concurrently, producing a dual AGN. This dual AGN population dwindles with decreasing redshift, as found in observations. Specific accretion rate and Eddington ratio distributions are in good agreement with observational estimates. The BH population is dominated in turn by fast, slow, and very slow accretors, with transitions  occurring at $z=3$ and $z=2$ respectively. 
\end{abstract}


\section{Introduction}\label{sec:Introduction}

Compelling evidence for massive black holes (BHs) exists for the nuclei of tens of nearby galaxies \citep[e.g.,][and references therein]{2013ARA&A..51..511K}. Scaling relations have been identified between the BH masses and the large-scale properties of the host galaxy, such as mass, luminosity, and velocity dispersion, pointing to a co-evolution between BHs and galaxies \citep[e.g.,][and references therein]{2011Sci...333..182H}. Active Galactic Nuclei (AGN), powered by accreting BHs, have been suggested to be the main ``regulators" of the BH growth itself, and also of the star formation rate in their host galaxies, at least in massive systems \citep[e.g.,][and references therein]{2011IAUS..277..273S}.  The relationship between galaxies and BHs has been suggested to hold the key to understanding massive galaxy evolution. 

AGN are also ``lighthouses" at early cosmic times: through multi-wavelength campaigns we can track the AGN population and its evolution \citep[e.g.,][and references therein]{Hop_bol_2007,2008ApJ...679..118S,2013ApJ...773...14R}, and use it as a signpost for the underlying BH population \citep[e.g.,][]{Merloni2004,Marconi2004,Shankar2004,Hopkins2006,Merloni08,2012ApJ...751...72D}. Through a wealth of observations we know that BHs are part of the evolution of cosmic structures, but how they formed, grew and interacted with their host galaxies is still unclear. 

The joint evolution of BHs and galaxies within the evolving cosmos can be studied, theoretically, through different techniques: analytical, semi-analytical and numerical.  We recall here the seminal papers by e.g., \cite{1992MNRAS.259..725S}, \cite{Silk1998}, \cite{HNR1998}, \cite{Haiman2000}, \cite{2000ApJ...543..599C} that first suggested, through analytical arguments, how BHs evolve cosmologically. Later on, semi-analytical models have been successful in matching the evolution of, e.g., the luminosity function of quasars \citep{Kauffmann2000,Cattaneo2001,VHM} and the scaling between BH mass and galaxy properties \citep{2000MNRAS.311..279M,2000MNRAS.318L..35H,Cattaneo2005} and showed how the inclusion of AGN feedback improved the match between predicted and observed galaxy mass and luminosity functions \citep{Croton2006, 2006MNRAS.370..645B,2007MNRAS.375.1189M,Hirschmann2012}. We refer the reader to \cite{Fontanot2011} for a review of the results of semi-analytical models pertaining to BHs and AGN. 

More recently, hydrodynamical cosmological simulations have also included BH evolution. Compared to semi-analytical models, a self consistent simulation is a quantitative asset  in that  semi-analytical models do not capture the anisotropic gas accretion and how galaxies interact while merging.   Many studies \citep{Sijacki2007, Dimatteo2008, Booth2009,duboisetal10,2012MNRAS.420.2662D,2014MNRAS.442.2304H,2015MNRAS.452..575S} have shown that the inclusion of BHs as ``sink particles" that are allowed to accrete gas from their surroundings, and produce in turn energy, a fraction of which is released in the host halo, reproduce the global evolution of the BH+galaxy population. This can appear surprising, as cosmological simulations are far from being able to resolve the scales over which accretion on BHs take place. Nevertheless, the {\it statistical} properties of AGN and BHs are sufficiently well reproduced. This may perhaps reflect that most of the growth of the BHs and the effects on their surroundings of AGN feedback are linked to short and intense phases, as previously noted in semi-analytical studies. 

In this paper we analyse Horizon-AGN\footnote{http://www.horizon-simulation.org/about.html} \citep{2014MNRAS.444.1453D}, a large-volume cosmological hydrodynamical simulation, with a physical resolution of $\sim 1$ kpc, run with the adaptive mesh refinement (AMR) code {\sc ramses}~\citep{teyssier02}. Compared to most existing studies \citep[e.g.,][]{Dimatteo2008, Booth2009,2015MNRAS.452..575S}, there are three aspects where the {\sc ramses} implementation differs  \citep[but see][]{Bellovary10,Tremmel}. These are: (i) BHs are not placed in halos only based on a halo mass threshold, but on the properties of gas in galaxies; (ii) BH are not anchored to the centre of the host or advected when far from the centre, (iii) BH feedback has a dual mode, thermal at high accretion rates and through a collimated jet at low accretion rates. These more realistic physical implementations allow us to push forward in the comparison between theoretical models and observations, and make additional predictions on the link between BHs and their hosts.

A series of papers  \citep{kavirajetal2015,duboisetal2015,peiranietal2015} have analysed the galaxy population in Horizon-AGN, and showed a good agreement with several observables (mass/luminosity functions, morphological mix, galaxy sizes, etc.). In this paper we validate the BH and AGN population and the BH-galaxy  cross-correlated properties (BH-galaxy scaling relations). A resolution study for this implementation has been done in Dubois et al. 2012, from 0.5 to 4 kpc varying the spatial resolution, and dark matter mass resolution as well. BH-galaxy scaling relations were similar in slope and normalisation. However, at low-masses, the BH population was affected, and the reader should keep this in mind; we also stress in the body of the paper the instances where Horizon-AGN is  affected by resolution issues. We then identify the limits where observables are well reproduced and further propose additional predictions and observational comparisons. 

The outline of the paper is as follows. In section 2 we briefly review the BH and AGN implementation. In section 3 we validate the simulation against several observational constraints. In section 4 we consider the connection between BHs, galaxies and dark matter halos, and instances where tidal stripping breaks the BH-host relations. In section 5 we discuss galaxies that host no BH, or more than one BH or AGN. In section 6 we follow the ups and downs of BH growth over cosmic time, analysing the distribution of accretion rates (in Eddington units, or per unit stellar mass), the duty cycle, and highlight how BHs go through different modes of accretion as a function of cosmic time. In section 7 we summarise our results and conclude.

\section{The Horizon-AGN simulation}\label{sec:H-AGN}

The Horizon-AGN simulation is described in detail in other papers, e.g., \cite{2014MNRAS.444.1453D}. We recall here the information relevant for BH physics, and for the analysis we performed. 

The Horizon-AGN simulation is run with the Adaptive Mesh Refinement code {\sc ramses}~\citep{teyssier02}. We adopt a standard $\Lambda$CDM cosmology with total matter density $\Omega_{\rm m}=0.272$, dark energy density $\Omega_\Lambda=0.728$, amplitude of the matter power spectrum $\sigma_8=0.81$, baryon density $\Omega_{\rm b}=0.045$, Hubble constant $H_0=70.4 \, \rm km.s^{-1}.Mpc^{-1}$, $n_s=0.967$ compatible with WMAP-7 cosmology~\citep{komatsuetal11}. The size of the box is $L_{\rm box}=100 \, h^{-1} {\rm Mpc}$ with $1024^3$ DM particles, which gives a dark matter (DM) mass resolution of $M_{\rm DM, res}=8\times 10^7 \, \rm M_\odot$.
From the level 10 coarse grid, a cell is refined (or unrefined) up to an effective resolution of $\Delta x=1$ proper kpc (level 17 at $z=0$) when the mass in a cell is more (or less) than 8 times that of the initial mass resolution.
The simulation includes prescriptions, described in more details in~\cite{2014MNRAS.444.1453D}, for background UV heating, gas cooling including the contribution from metals released by stellar feedback, star formation following a Schmidt law with a 1 per cent efficiency~\citep{rasera&teyssier06}, and feedback from stellar winds and type Ia and type II supernovae (SNae) assuming a Salpeter initial mass function~\citep{dubois&teyssier08,KimmPhD}.

BHs are created in cells where the gas and stellar density density exceeds the threshold for star formation, $n_0=0.1\, \rm H\, cm^{-3}$, and where the stellar velocity dispersion, in that cell, is larger than $100\, \rm km\,s^{-1}$, with an initial seed mass of $10^5\, \rm M_\odot$. In order to avoid the formation of multiple BHs in the same galaxy, BHs are not allowed to form at distances smaller than 50 comoving kpc from any other BH particle. BH formation is stopped at $z=1.5$.  All galaxies with mass $>10^{10} \msun$ at $z=0$ have already formed by that point, i.e., they have at least one progenitor where a BH would have been seeded. 

We use the same ``canonical'' accretion and AGN feedback modelling employed in~\cite{2012MNRAS.420.2662D}. The accretion rate onto BHs follows the Bondi-Hoyle-Lyttleton rate $\dot M_{\rm BH}=4\pi \alpha G^2 M_{\rm BH}^2 \bar \rho / (\bar c_s^2+\bar u^2) ^{3/2},$
where $M_{\rm BH}$ is the BH mass, $\bar \rho$ is the average gas density, $\bar c_s$ is the average sound speed, $\bar u$ is the average gas velocity relative to the BH velocity, and $\alpha$ is a dimensionless boost factor with $\alpha=(\rho/\rho_0)^2$ when $\rho>\rho_0$ and $\alpha=1$ otherwise~\citep{Booth2009} in order to account for our inability to capture the colder and higher density regions of the ISM.
The effective accretion rate onto BHs is capped at the Eddington luminosity with an assumed radiative efficiency of $\epsilon_{\rm r}=0.1$ for the \cite{Shakura73} accretion onto a Schwarzschild BH.

In order to avoid spurious oscillations of the BH in the gravitational potential well due to external perturbations and finite resolution effects, we introduce a drag force that mimics the dynamical friction exerted by the gas onto a massive particle. 
This dynamical friction is proportional to $F_{\rm DF}=f_{\rm gas} 4 \pi \alpha \rho (G M_{\rm BH}/\bar c_s)^2$, where $f_{\rm gas}$ is a fudge factor whose value is between 0 and 2 and is a function of the mach number ${\mathcal M}=\bar u/\bar c_s<1$~\citep{ostriker99, chaponetal13}, and where we introduce the boost factor $\alpha$ for the same reasons than stated above. See also discussions in \cite{2015MNRAS.451.1868T,2015MNRAS.446.1765L,2015arXiv150802224G} on the limitations of gravity solvers for BH dynamics in AMR and SPH simulations. 

The AGN feedback is a combination of two different modes, the so-called \emph{radio} mode operating when $\chi=\dot M_{\rm BH}/\dot M_{\rm Edd}< 0.01$ and the \emph{quasar} mode active otherwise.
The quasar mode corresponds to an isotropic injection of thermal energy into the gas within a sphere of radius $\Delta x$, at an energy deposition rate: $\dot E_{\rm AGN}=\epsilon_{\rm f} \epsilon_{\rm r} \dot M_{\rm BH}c^2$,
where $\epsilon_{\rm f}=0.15$ for the quasar mode is a free parameter chosen to reproduce the correlations between BHs and galaxies and the BH density in our local Universe (see \citealp{2012MNRAS.420.2662D}).
At low accretion rates, on the other hand, the radio mode deposits the AGN feedback energy into a bipolar outflow with a jet velocity of $10^4\,\rm km\, s^{-1}$ into a cylinder with a cross-section of radius $\Delta x$ and height $2 \, \Delta x$ following~\cite{ommaetal04} (more details about the jet implementation are given in~\citealp{duboisetal10}).
The efficiency of the radio mode is larger, with $\epsilon_{\rm f}=1$.

We identify dark matter halos and sub-halos using HaloMaker, which uses AdaptaHOP \citep{Aubert+04,Tweed+09}, a structure finder based on the identification of saddle points in the (smoothed) density field.  A total of 20 neighbours are used to compute the local density of each particle, and we fix the density threshold at 178 times the average total matter density. The force softening (minimum size below which substructures are considered irrelevant) is $\sim$ 2 kpc. Only dark matter halos identified with more than 50 particles are considered. Galaxies are identified using the same method and same parameters but using the stellar particle distribution instead of the dark matter one. Because of AGN feedback, dark matter density profiles flatten in the centre, and misidentification of the halo centre can occur when it is purely determined by the densest particle in the halo (Peirani et al. in prep.): a small cusped sub-halo can be identified as the centre of the main more massive cored halo.
We use the shrinking sphere approach proposed by~\cite{poweretal03} to get the correct halo centre. From a first guess of the centre using the densest particle, we recursively search the centre of mass within a radius 10\% smaller than the radius at the previous iteration (starting  from the virial radius). We end  the search when the sphere has a size smaller than 2 kpc, and take the densest particle within that sphere as the new halo centre.

\begin{figure*}
\centering
\includegraphics[width=0.75\textwidth,angle=0]{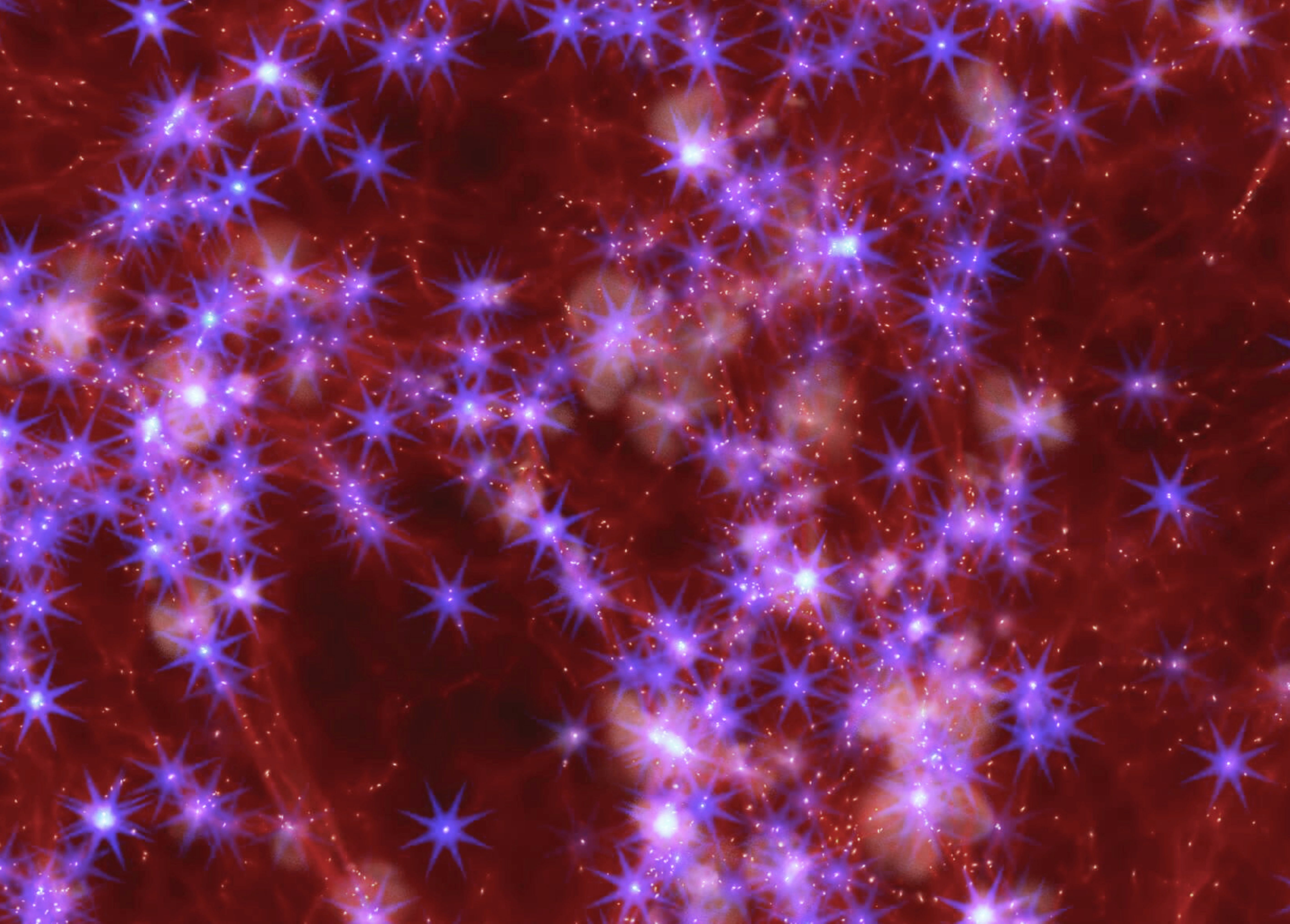}
\includegraphics[width=0.75\textwidth,angle=0]{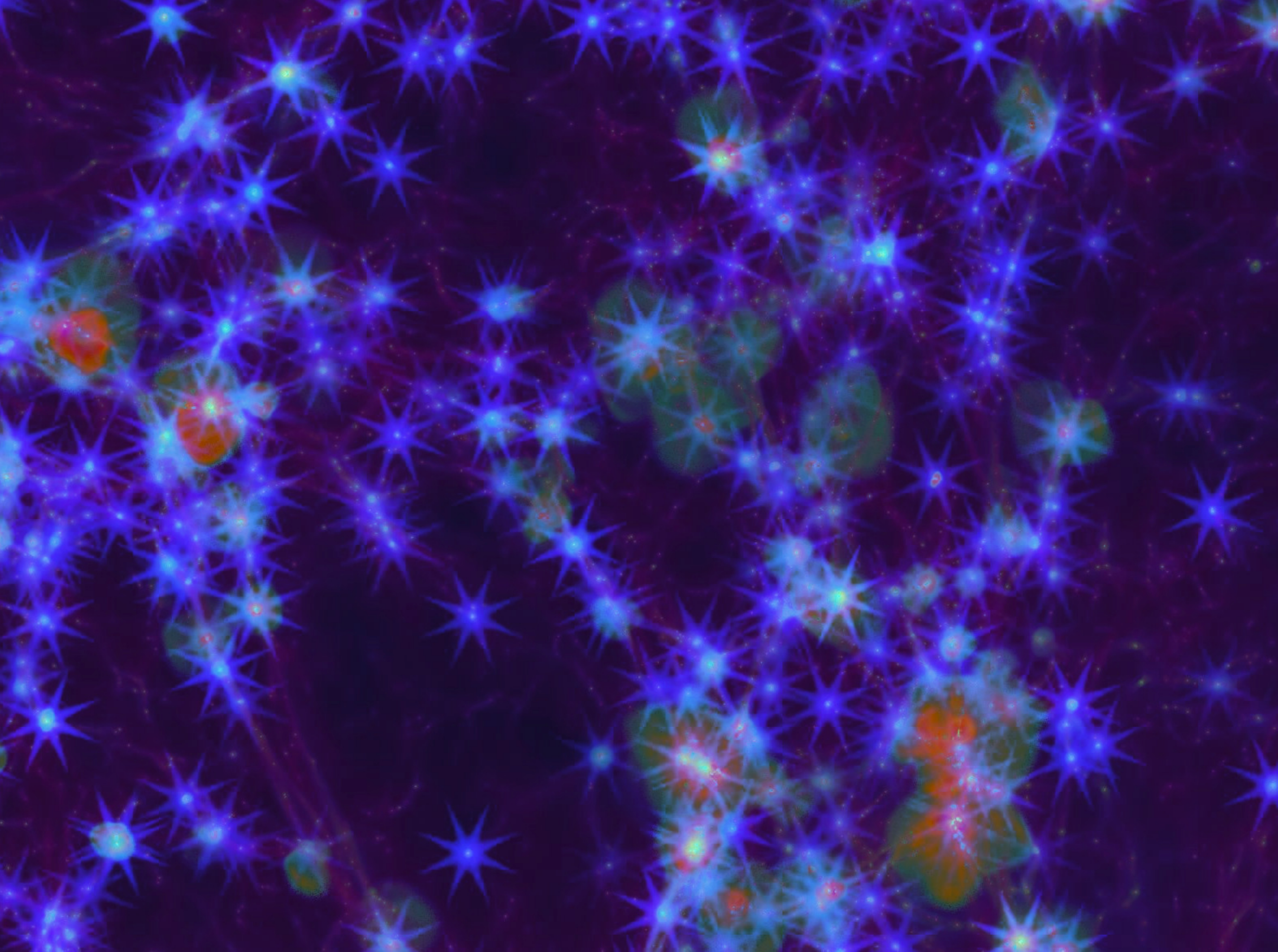}
\caption{The positions of BHs, shown as diffraction spikes, in a slice of 200$\times$142$\times$50 Mpc/h are overlaid on a gas density and metallicity (top) or density and temperature (bottom) map in Horizon-AGN at $z\sim 1.5$.}
\label{fig:map}
\end{figure*}

In Horizon-AGN BHs are not anchored in the centre of dark matter halos, therefore we have to assign BHs to their host halos and galaxies. We assign a BH to a halo if it is within 10 per cent of the virial radius of the dark matter halo and to a halo+galaxy structure if the BH is also within twice the effective radius of the most massive galaxy within that halo. If more than one BH meets the criteria, the most massive BH is defined as the central BH, and removed from the list.  We start from main halos and proceed hierarchically through sub-halos. Some BHs are not assigned to any halo, subhalo, or galaxy. These are treated as ``off-centre" BHs and described in section~5. In Fig.~\ref{fig:map} we show a map of gas density with BH positions overlaid.

\section{Comparison with observations}
Our first step is to validate our simulation, i.e., to demonstrate that it produces a population of BHs compatible with the real Universe. For this check, we consider four main diagnostics: the luminosity function of AGN, the BH mass function, the BH mass density versus redshift and the relation between BH and galaxy properties. 

\subsection{AGN luminosity function}

\begin{figure}
\centering
\includegraphics[width=\columnwidth,angle=0]{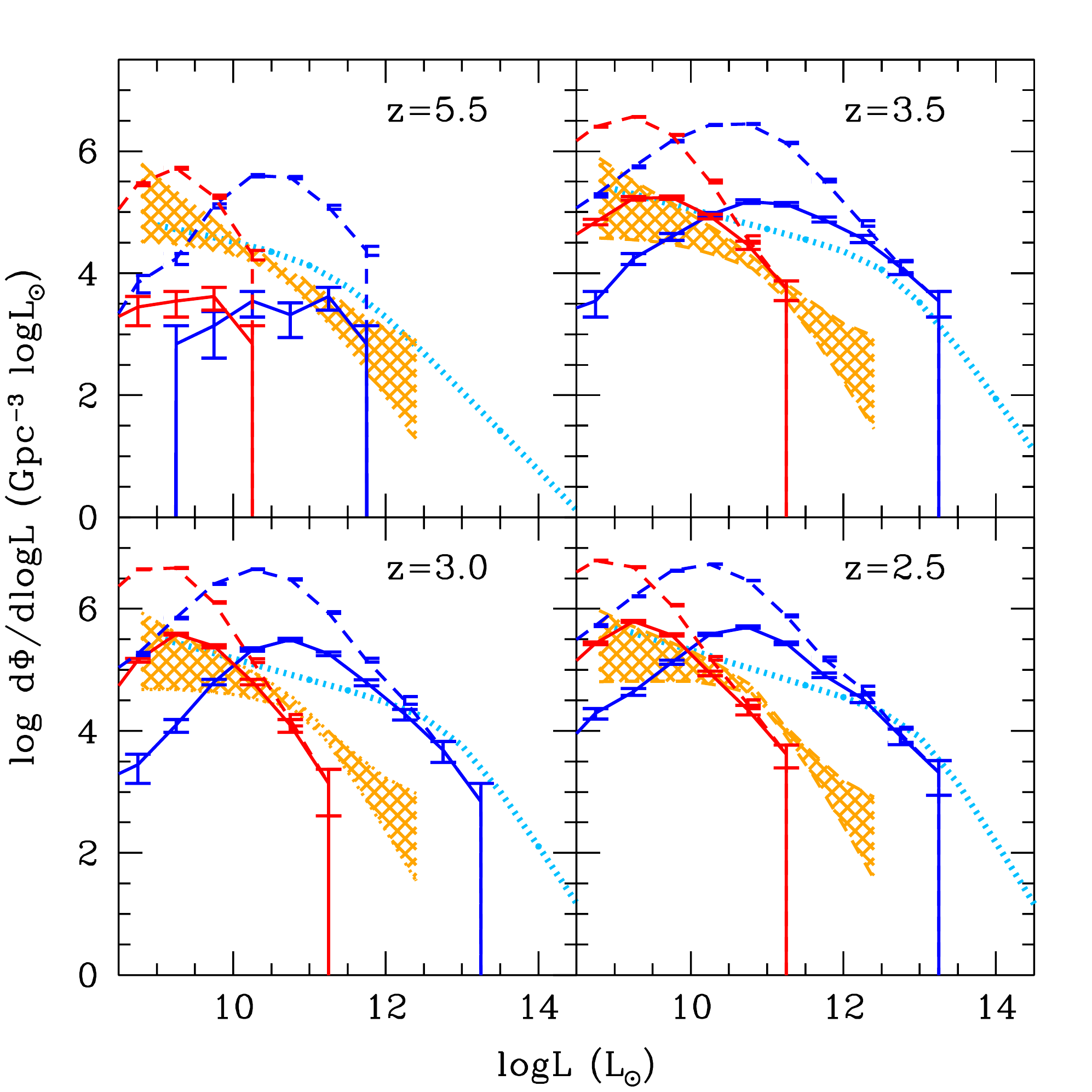}
\includegraphics[width=\columnwidth,angle=0]{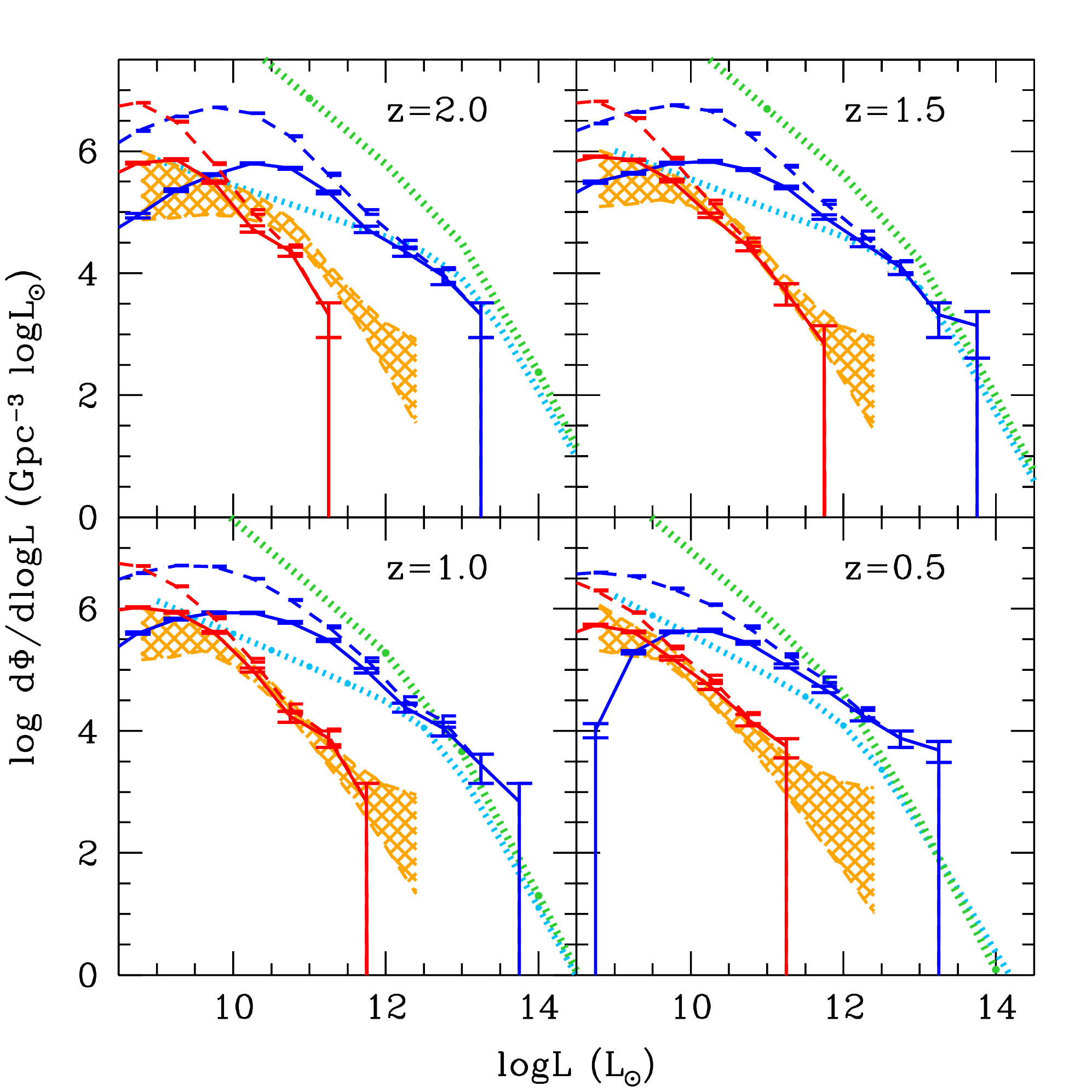}
\caption{Bolometric (blue) and hard X-ray (2-10 keV, red) LF at different redshifts. The dashed curves include all BHs in Horizon-AGN in halos with mass $>8\times 10^{10}\msun$, while the solid curves include only BHs in halos with mass $>5\times 10^{11}\msun$, where SN feedback should not quench BH growth \citep{2015MNRAS.452.1502D}. The dotted light blue curve is the bolometric LF proposed by \citet{Hop_bol_2007} (see also \citealt{2009ApJ...690...20S} and \citealt{2014ApJ...786..104U}). The orange hatched region is the hard X-ray LF by \citet{2015ApJ...802...89B} corrected for Compton thin and thick AGN. The green dotted curve is the luminosity function derived from a spectroscopic survey of AGN selected from Spitzer Space Telescope imaging surveys \citep{2015ApJ...802..102L}.}
\label{fig:LF}
\end{figure}

We compare the theoretical luminosity function (LF, $\Phi$) to the observational determination of the bolometric LF, of which different  estimates exist \citep{Hop_bol_2007, 2009ApJ...690...20S, 2014ApJ...786..104U}. We choose here the functional form proposed by \cite{Hop_bol_2007}, which at the redshifts of interest falls between those proposed by  \cite{2014ApJ...786..104U} and \cite{2009ApJ...690...20S}, thus marking a ``middle ground".  We also adopt the most recent hard X-ray LF \citep[2-10 keV,][]{2015ApJ...802...89B}, which includes a correction for obscured ($10^{22}<N_H<10^{24}\, {\rm cm}^{-2}$, where $N_H$ is the equivalent neutral hydrogen column density) and Compton Thick AGN ($N_H>10^{24}\, {\rm cm^{-2}}$) -- but estimating the contribution of the elusive Compton Thick AGN remains challenging.  \cite{2016A&A...587A.142F}, who estimate the LF at [5-10 keV], find no evidence of a rapid decline of the LF up to redshift four, consistently with \citet{2015ApJ...802...89B}.   For this comparison we convert bolometric luminosities to X-ray using the bolometric correction from \cite{Hop_bol_2007} for consistency. To give an idea of the current uncertainties, we also include, at $z\lesssim2$, the conversion to bolometric LF of the mid-infrared selected AGN LF proposed by \cite{2015ApJ...802..102L}, where they suggest a larger population of faint AGN exists, that can be missed at other wavelengths. 

The resulting LFs at different redshifts are shown in Fig.~\ref{fig:LF}. The bright end of the LF is generally in good agreement with observations, while the faint end is overestimated when including BHs in all halos above the nominal resolution of $8\times 10^{10}\msun$, i.e., resolved by more than 1000 DM particles (dashed curves). The overestimate of the faint end of the LF is a common problem in cosmological simulations, for instance \cite{2015MNRAS.452..575S} find a similar behaviour in the Illustris simulations. They suggest that a good agreement is reached when including only BHs with mass $>5\times 10^7\, \msun$, and discuss other sources of possible discrepancy with data. We suggest that the main source of the overestimate is due to SN feedback. \cite{2015MNRAS.452.1502D} find that strong SN feedback (using a ``delayed cooling" prescription from~\citealp{teyssieretal13}) in low-mass galaxies (or halos) is able to limit BH growth by SN-driven outflows, which deplete the central region of the galaxy of gas. As the galaxy and its halo accumulate mass, they become able to confine nuclear inflows and the BH can start to grow. If we included only BHs with halo mass $>5\times 10^{11}\msun$ (solid curves), where BHs growth is unimpeded by SN feedback, the agreement greatly improves. Similar results are obtained with a cut in galaxy mass $>3\times 10^{10}\msun$ at the threshold where, in their higher-resolution simulations with delayed cooling SN feedback,  \cite{2015MNRAS.452.1502D} find that BHs become unaffected by SN feedback. Using a minimum threshold BH mass of $>2-3\times 10^{7}\msun$ leads to analogous results for the luminosity function. A direct comparison of BH growth under the effect of SN feedback in a cosmological volume is subject of a paper in preparation (Habouzit et al.), but we stress that the mass locked in BHs hosted in low-mass halos, $<5\times 10^{11}\msun$, is relatively unimportant from the cosmological point of view, as shown in the following section.  Additionally, simulations may overproduce the high redshift AGN LF, Êbut given the rather poor observational constraints at high-$z$ and uncertainties in the spectral energy distribution, the discrepancy may eventually be reduced or disappear when better and more calibrated high-$z$ LFs become available. There is also a possible overestimate of the high-luminosity end at $z<1.5$, but the statistics are poor, and this is seen only in the bolometric LF, not in the hard X-ray LF.

\subsection{BH mass density and mass function}

\begin{figure}
\centering
\includegraphics[width=\columnwidth,angle=0]{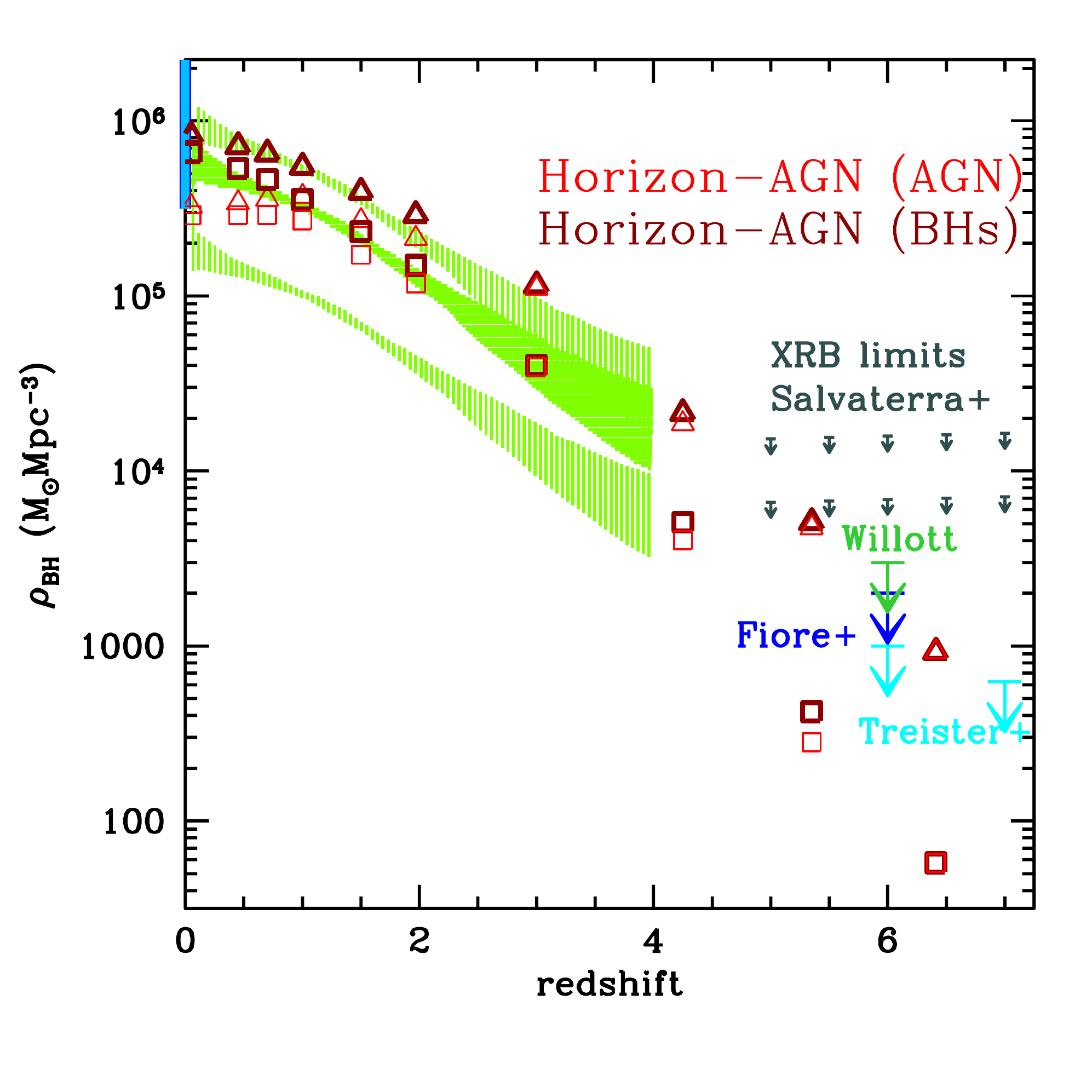}
\caption{Total mass density locked in BHs (dark red), and BH mass density in active BHs (light red), defined as those with bolometric luminosity $>10^{43} {\rm erg\, s^{-1}}$ versus redshift.  Triangles: BHs in halos with mass $>8\times 10^{10}\msun$, squares: BHs in halos with mass $>5\times 10^{11}\msun$ (see text for details). The green hatched region shows the limits provided by Soltan's argument, as proposed by Buchner et al. (2015) using a radiative efficiency of 0.1, and scaling the radiative efficiency between 0.06 and 0.32 (upper and lower hatched regions respectively). The vertical blue line indicates the $z=0$ BH mass density estimated by Shankar et al. (2004), and correcting the BH-to-bulge relation as suggested by Kormendy \& Ho (2013). The limits at $z>6$ are derived from stacked high-z galaxies (Willott et al. 2011; Fiore et al. 2012; Cowie et al. 2012, Treister et al. 2013) or from the integrated X-ray background (Salvaterra et al. 2012); Compton thick sources are not included in these limits.}
\label{fig:rho}
\end{figure}

The BH mass density, and its evolution over cosmic time, are important observational diagnostics. The way they are calculated observationally is worth a discussion prior to the comparison with the theoretical results from Horizon-AGN. 

\begin{figure}
\centering
\includegraphics[width=\columnwidth,angle=0]{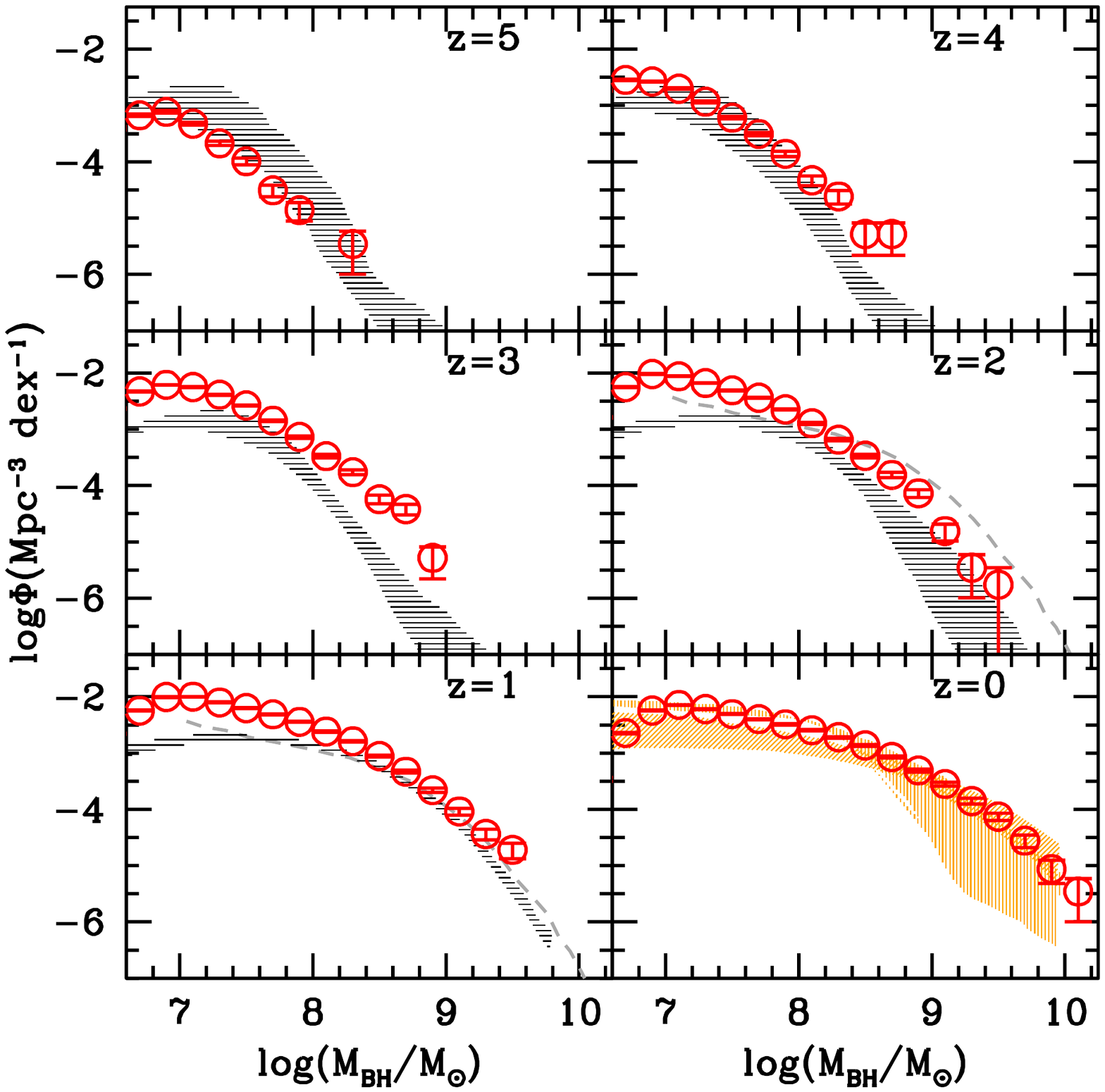}
\caption{Mass function of BHs in Horizon-AGN at different redshifts.  We indicate with the hatched regions the range proposed by estimates at $z=0$ \citep[vertical and diagonal hatching respectively]{Lauer2007,2013CQGra..30x4001S}, and, at high redshift, the range proposed at $z=1.5$ by \citet{2015MNRAS.447.2085S}, shown as grey dashed curves at $z=1$ and $z=2$,  and by \citet{Merloni08} at $z=1-5$, shown with horizontal black hatching.}
\label{fig:MF}
\end{figure}

At $z=0$ the {\it total BH mass density}, $\rho_{\rm BH}$, is estimated starting from the empirical correlations between BH mass and galaxy properties (normally, the velocity dispersion, the bulge mass and luminosity). The BH mass density is calculated through the convolution of the chosen correlation with the distribution function of galaxies as a function of that property \citep{YuTremaine2002}. This means, for instance, the galaxy mass function and luminosity functions (corrected for the bulge fraction) and the probability distribution function for the velocity dispersion \citep[and references therein]{2013CQGra..30x4001S}. In principle, this BH mass density is driven by the correlations found on the local sample of {\it quiescent} BHs, less than $150$ BHs at the time of writing, then extrapolated to the global population, including AGN \citep[see,][and references therein, for a discussion on the case of active BHs]{2015ApJ...813...82R}. The shape, normalization and scatter in the relationships play an important role. For instance, \cite{2013ARA&A..51..511K} suggest that the ratio between BH and bulge mass previously used in benchmark calculations \citep[e.g.,][]{Shankar2004,Marconi2004} should be increased, based on their updated sample, leading to a total mass density higher by a factor of almost 2.

At $z>0$ the shape and normalization of the BH-galaxy correlations are subject of debate \citep[e.g.,][]{2006ApJ...649..616P,Jahnke2009,Merloni2010,2011ApJ...741L..11C,2012MNRAS.420.3621T,2014MNRAS.443.2077B}, and therefore large uncertainties exist. Typically, Soltan's argument \citep{Soltan1982} and its updates \citep{1999MNRAS.303L..34F,YuTremaine2002,Elvis2002,Hop_bol_2007,2016LNP...905..101M} are used instead to estimate the BH mass density. The luminosity function of AGN is integrated over time, starting from a given cosmic time $t_{\rm max}$ from which the LF is known sufficiently well, and rescaled by a (fixed) radiative efficiency, $\epsilon$, to obtain a mass density, $\rho_{\rm BH,acc}(z)\leqslant \rho_{\rm BH}(z)$, of {\it mass accreted on BHs} as a function of redshift:

\begin{equation}
\rho_{\rm BH,acc}(z)= \frac{1-\epsilon}{\epsilon c^2}\int_{t_{\rm max}}^{t(z)} dt \int dL\,L\,\Phi_{\rm AGN}(L,t).
\end{equation}

In general, the population of BHs in Horizon-AGN is in good agreement with a variety of observations based on Soltan's argument, from high redshift to $z=0$ (Fig.~\ref{fig:rho}), but we can also derive additional information and constraints.  The density of mass accreted on BHs is obviously a lower limit to the total mass density in BHs, which includes quiescent ones and the initial mass in ``seeds" \citep[see][]{2010A&ARv..18..279V,Volonteri2013,2016LNP...905..101M}, plus obscured AGN \citep{2016LNP...905..101M}.  This is exemplified in Fig.~\ref{fig:rho}, where we distinguish the total BH mass density (dark red) and the mass density in accreting BHs in the AGN phase (bolometric luminosity $>10^{43} {\rm erg\, s^{-1}}$, light red). The latter quantity is the true mass density in accreting BHs at a given redshift, rather than the total luminosity rescaled by the radiative efficiency ($\epsilon=0.1$ in Horizon-AGN) and the speed of light, integrated over time. This quantity allows us to immediately see the relevance of the bright AGN phases in growing BHs. At high redshift, $z>2$, most of the BH mass  is grown in luminous AGN, while at lower redshift the mass growth is driven by  faint AGN (bolometric luminosity $<10^{43} {\rm erg\, s^{-1}}$). This figure also shows that BHs in low-mass halos ($<5\times 10^{11}\msun$) are relatively unimportant in the cosmic picture as one can see by comparing triangles and squares, which include BHs in halos with mass $>8\times 10^{10}\msun$ or  $>5\times 10^{11}\msun$ respectively. Finally, the total mass density in seeds is $\lesssim 10^3 \msun {\rm Mpc}^{-3}$, similar to what is predicted by analytical and semi-analytical models of BH formation \citep{2010A&ARv..18..279V}, which means that accretion accounted for basically the whole mass density we find at $z=0$, corresponding to $\sim 8.3\times 10^5 \msun {\rm Mpc}^{-3}$.

Horizon-AGN reproduces well the BH mass density, but this does not necessarily ensure that it reproduces the BH mass function. The BH mass density is  the integral of the BH mass function, and it is dominated by BHs at its knee, therefore reproducing the mass density does not give information on low- and high-mass BHs. In Fig.~\ref{fig:MF} we show the BH mass function  (red dots), and attempt a comparison with observational estimates, both at $z=0$ and at high redshift. The comparison is complicated by trying to match the theoretical sample to observations. We refer the reader to the comprehensive review by \cite{2012AdAst2012E...7K} for a discussion on the difficulties in determining the BH mass function observationally, and the related uncertainties.  We discuss in the following the main problems of relevance for our comparison here. Regarding the comparison at $z=0$, we use  the results from \cite{Lauer2007} and \cite{2013CQGra..30x4001S} \citep[see also][]{Shankar2004}. As noted previously for the mass density, the estimates of the BH mass function hinge on a convolution of the correlations between BH and galaxy properties with the distribution function of galaxies as a function of that property. The functional form and the range of validity of these correlations are subject of debate, and therefore a source of uncertainty. For instance, the slope of the BH mass-velocity dispersion correlation, and the normalization of the BH-bulge mass one are still uncertain. At higher redshift, the mass function has been estimated only for  broad-line quasars and AGN \citep[Type 1,][]{2010ApJ...719.1315K,2013ApJ...764...45K,2015MNRAS.447.2085S}, which are only the subset of quasars for which broad lines can be seen, i.e., with a favourable viewing angle and unobscured (obscuration may come from the geometry of the torus, and depend only on viewing angle, but also by material in the host galaxy, which is harder to account for).  We include  in Fig.~\ref{fig:MF}, at $z=1$ and $z=2$, the estimate by \cite{2015MNRAS.447.2085S} at $z=1.5$ based on a convolution of the observational stellar mass function \citep{2013A&A...556A..55I} with the BH-bulge mass relation of \cite{McConnell2012a}, without applying any bulge-total mass conversion. We also include the mass function proposed by \cite{Merloni08} obtained by modelling the BH population through a continuity equation derived from using the redshift evolution of the LF and the BH mass function at $z=0$ as constraints.  The agreement between the simulation and the observations is good; the disagreement at low BH masses ($<3\times 10^7 \msun$) at $z=1-3$ is similar  to what we find for the AGN LF. It is useful to contrast the BH mass function to the stellar mass function in horizon-AGN (Kaviraj et al. in prep.). The simulated galaxies reproduce well the observational stellar mass function at $z=0-5$, with only an overestimate of $\sim 0.3$ dex for stellar masses $<3\times 10^{10} \msun$. This overestimate is somewhat smaller than the one we find in the BH mass function at $z=1-3$ at BH masses $<3\times 10^{7} \msun$.

From Fig.~\ref{fig:MF} we can also address the question of down to which BH mass the simulation is reliable.  The closer the BH mass is to the resolution, the more results are resolution dependent.  One approach is to proceed like in the case of galaxies and derive the minimal mass from the completeness of the BH mass function. The argument is as follows: there is no explicit scale breaking in our BH model, therefore there is no reason for the number of BHs not to increase with the number of host galaxies/halos. Consequently, any decrease in the mass function of BH at the low mass end could be attributed to lack of resolution in the simulation. Applying this argument the minimal BH mass is $2\times 10^7 \msun$. However, while there is no explicit scale breaking in our BH model, there is an implicit threshold on the minimum gas density required for BH seeding, which disfavours low-mass galaxies. Therefore, it does not necessarily follow that we should not trust BH with masses $ < 10^7 \msun$, this is saying that this sample is not complete in the simulation so our conclusions might be biased in that regime. We further discuss the range of validity of the simulation in the next section.

\begin{figure}
\centering
\includegraphics[width=\columnwidth,angle=0]{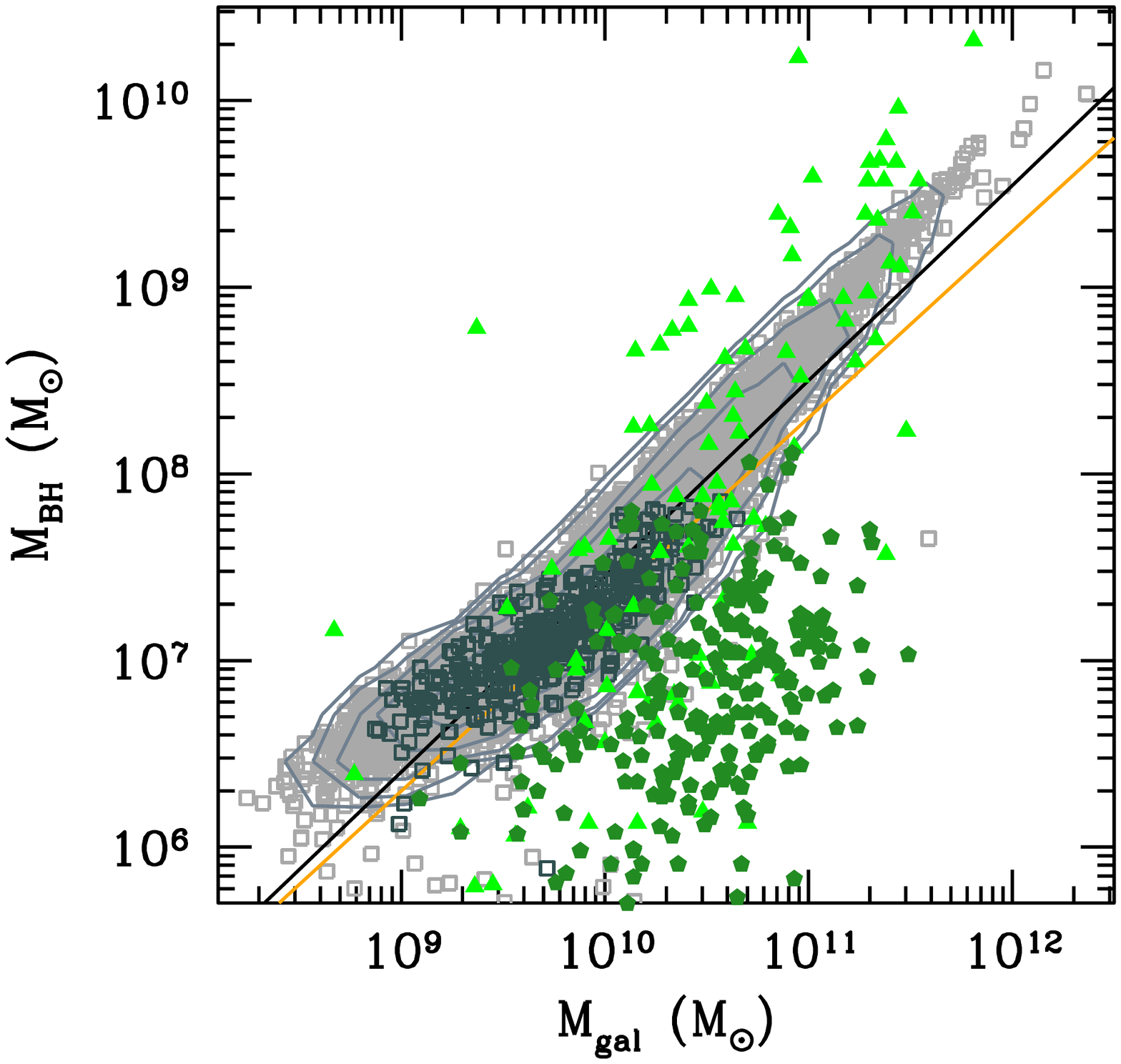}
\caption{BH versus  total stellar mass of the galaxy for 10,000 random galaxies at $z=0$ (grey).  Dark grey squares: BHs with  $L_{\rm bol}<10^{44} {\rm erg\, s^{-1}}$ and Eddington ratio $>0.01$. The orange line marks $M_{\rm BH}=2\times10^{-3} M_{\rm gal}$. The black line is the best fit linear correlation. The darker green pentagons are low luminosity AGN from Reines \& Volonteri (2015), and the lighter green triangles are the quiescent BHs from Tables 2 and 3 in Kormendy \& Ho 2013 with stellar masses revised by Reines \& Volonteri. The stellar masses of Horizon-AGN galaxies have been shifted by $-$0.33 dex to account for this correction.  If we select for low-luminosity AGN, their BHs occupy a region in the lower-right corner of the BH-galaxy distribution, reminiscent of the region where in observations we tend to find low-luminosity AGN. However, the scatter in the simulated sample is much less than in the observations, see text for details.}
\label{fig:BH_gal}
\end{figure}

\subsection{Correlations between BHs and galaxies at $z\sim 0$}

One of the most common ways to determine whether simulated BHs are realistic is to compare their masses to properties of the host galaxies that have been shown to correlate with BH masses. Indeed, in most cases the parameters of BH accretion and feedback are tuned to reproduce these correlations \citep[e.g.,][]{Sijacki2007,Dimatteo2008,Booth2009,2012MNRAS.420.2662D}.  In reality, a ``fair" comparison is far from simple. On the one hand, the sample of galaxies with dynamical BH mass measurements is not an unbiased representation of the whole galaxy population. Most BH masses pertain to massive, dense elliptical galaxies, while disc galaxies and low velocity dispersion galaxies are under-represented, as well as low-mass galaxies \citep{2015ApJS..218...10V}, with the rare exception of megamasers \citep{2010ApJ...721...26G}. Additionally, the definition of the galaxy property of interest (e.g., bulge mass, velocity dispersion, galaxy mass) is not consistent in the literature, and the analyses are heterogeneous.

We start with the galaxy property that is easier to define and measure in both simulations and observations, the total stellar mass of the galaxy, although whether the total stellar mass is a good predictor of the BH mass, and a tight correlation can be defined is a matter of debate \citep{2013ARA&A..51..511K,2014ApJ...780...70L}. Recently, \cite{2015ApJ...813...82R} re-analyzed in as much consistent a way as possible, the sample of local BHs with dynamical mass measurements \citep{2013ARA&A..51..511K}, and also include  BHs in AGN with mass measured through reverberation mapping \citep{2015PASP..127...67B}, as well as a sample of broad-line AGNs with single-epoch BH mass estimates selected from the NASA-Sloan Atlas, which is based on the Sloan Digital Sky Survey Data Release 8 spectroscopic catalog \citep{aiharaetal2011}.  They found that the relation between BH and stellar mass defined by local moderate-luminosity AGN  in relatively low-mass galaxies (pentagons in Fig.~\ref{fig:BH_gal}) has a normalization that is lower by approximately an order of magnitude compared to the BH-bulge mass relation, and the to the relation obtained on massive, bulge-dominated galaxies (triangles in Fig.~\ref{fig:BH_gal}), which appears also to be steeper.

\begin{figure}
\centering
\includegraphics[width=\columnwidth,angle=0]{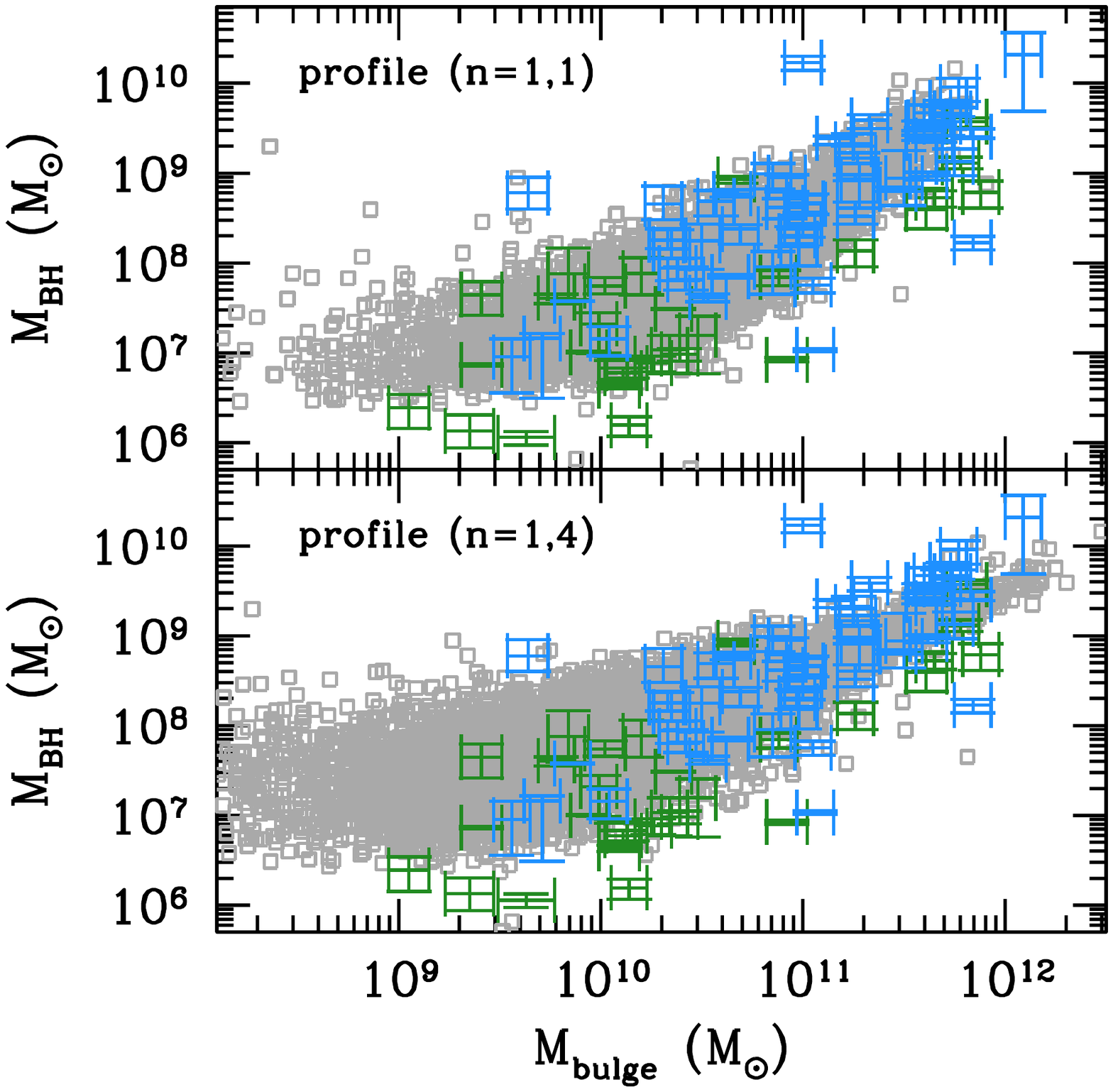}
\vspace{-2cm}
\includegraphics[width=\columnwidth,angle=0]{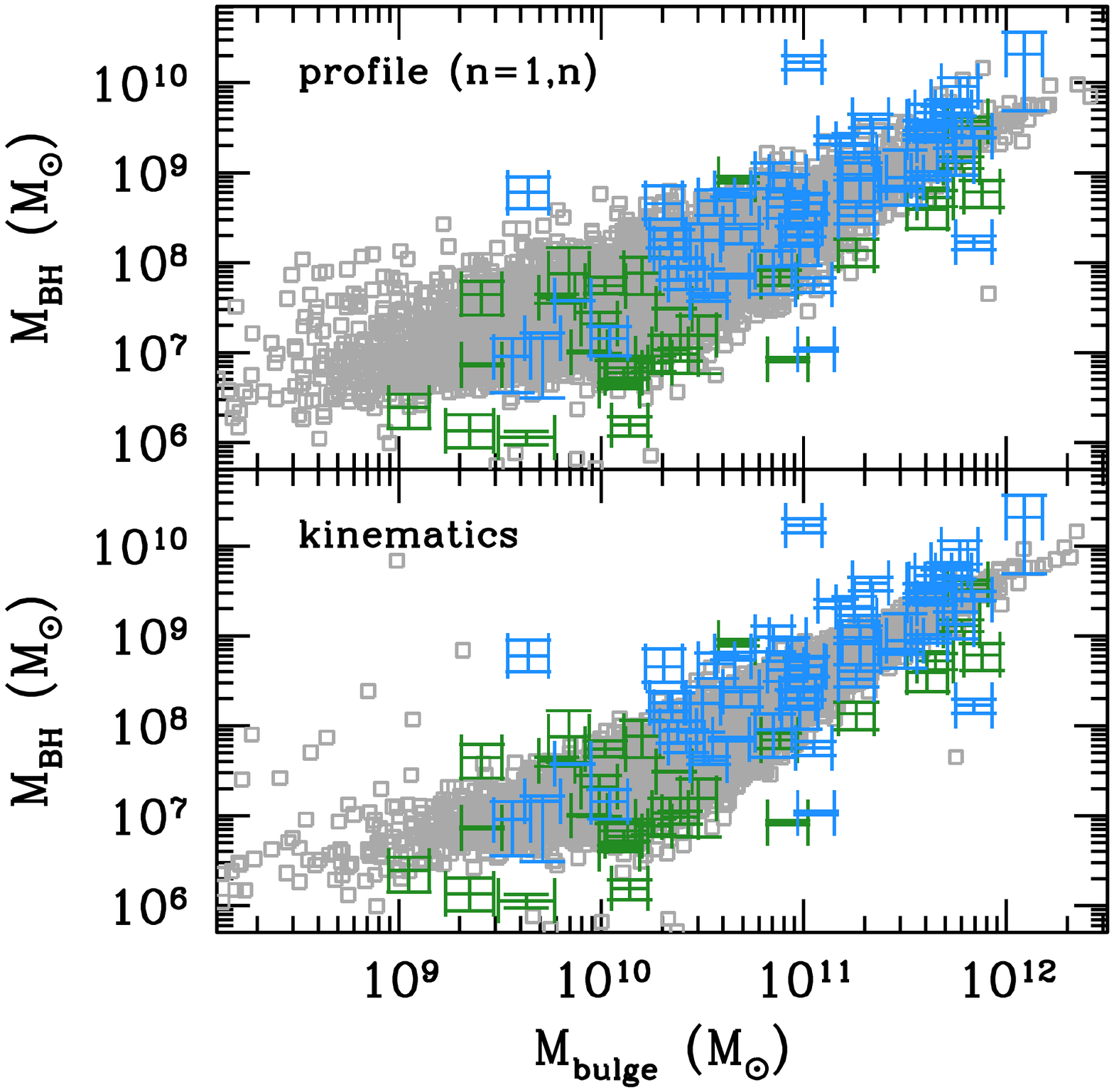}
\caption{BH versus bulge mass for 10,000 random galaxies at $z=0$ (grey). The bulge mass is measured either kinematically  or through a bulge/disc decomposition with two Sersic profiles, either both with $n=1$, or one with $n=1$, and one with $n=4$, or one with $n=1$, and one with $n=[1,4]$. We report also all BHs from Tables 2 and 3 in Kormendy \& Ho 2013, using blue symbols for the subset used for fitting the BH mass versus bulge mass relation.}
\label{fig:BH_bulge}
\end{figure}

 We define as galaxy mass the total stellar mass.  The stellar mass is the sum of all star particle masses within a region meeting a density threshold criterion, set at 178 times the average total matter density. Only galaxies identified with more than 50 particles are considered. (See section 2 for details.) Fig.~\ref{fig:BH_gal} reports 10,000 random BHs+galaxies from Horizon-AGN at $z=0$ (squares). We note that the BH parameters in {\sc ramses} had been originally chosen to reproduce older data and trends \citep{2012MNRAS.420.2662D}, therefore, to make the comparison with the observational sample self-consistent, the masses of  simulated galaxies have been shifted by $-$0.33 dex \cite[see the discussion in][]{2015ApJ...813...82R}.  At the high-mass end, $M_{\rm BH}>10^8 \, \msun$ and $M_{\rm gal}>10^{11}\, \msun$, simulations and observations are in relatively good agreement, but low-mass BHs and low-mass galaxies are more problematic. Specifically,  BHs  with $M_{\rm BH}<5\times 10^{7}\, \msun$ in intermediate mass galaxies ($10^{10}\, \msun<M_{\rm gal}<10^{11}\, \msun$) are almost absent. As discussed in section 3.1, this may be caused by the lack of strong SN feedback in Horizon-AGN, which allows BHs in low-mass galaxies to grow more easily.   SN feedback would affect only BHs in low-mass galaxies. Once a galaxy reaches a mass of $\sim 10^{10} \, \msun$ then the BH can grow unimpeded and self-regulation ensues, thus, for instance, in the simulations by \cite{2015MNRAS.452.1502D} the host galaxy properties did not depend on SN feedback after $z\sim 3$. In the presence of delayed cooling SN feedback the bulge growth was suppressed at $z>3$. Afterwards, when the galaxy mass reached a few $\times 10^{10} \, \msun$  the presence of the AGN regulated the bulge mass independently of the SN implementation. In analogy with this, we argue that, with smaller BHs, causing reduced AGN feedback in small and intermediate mass galaxies, the fraction of simulated bulge-dominated galaxies at stellar mass $\lesssim10^{11} \, \msun$ would decrease. This would improve the match with the observed morphological mix in that mass range (Dubois et al. in prep.). Additionally, as noted in section 3.2, below BH masses $2\times 10^7 \msun$ (consequently, galaxy mass $\lesssim 10^{10} \msun$) resolution effects become important.

\begin{figure}
\centering
\includegraphics[width=\columnwidth,angle=0]{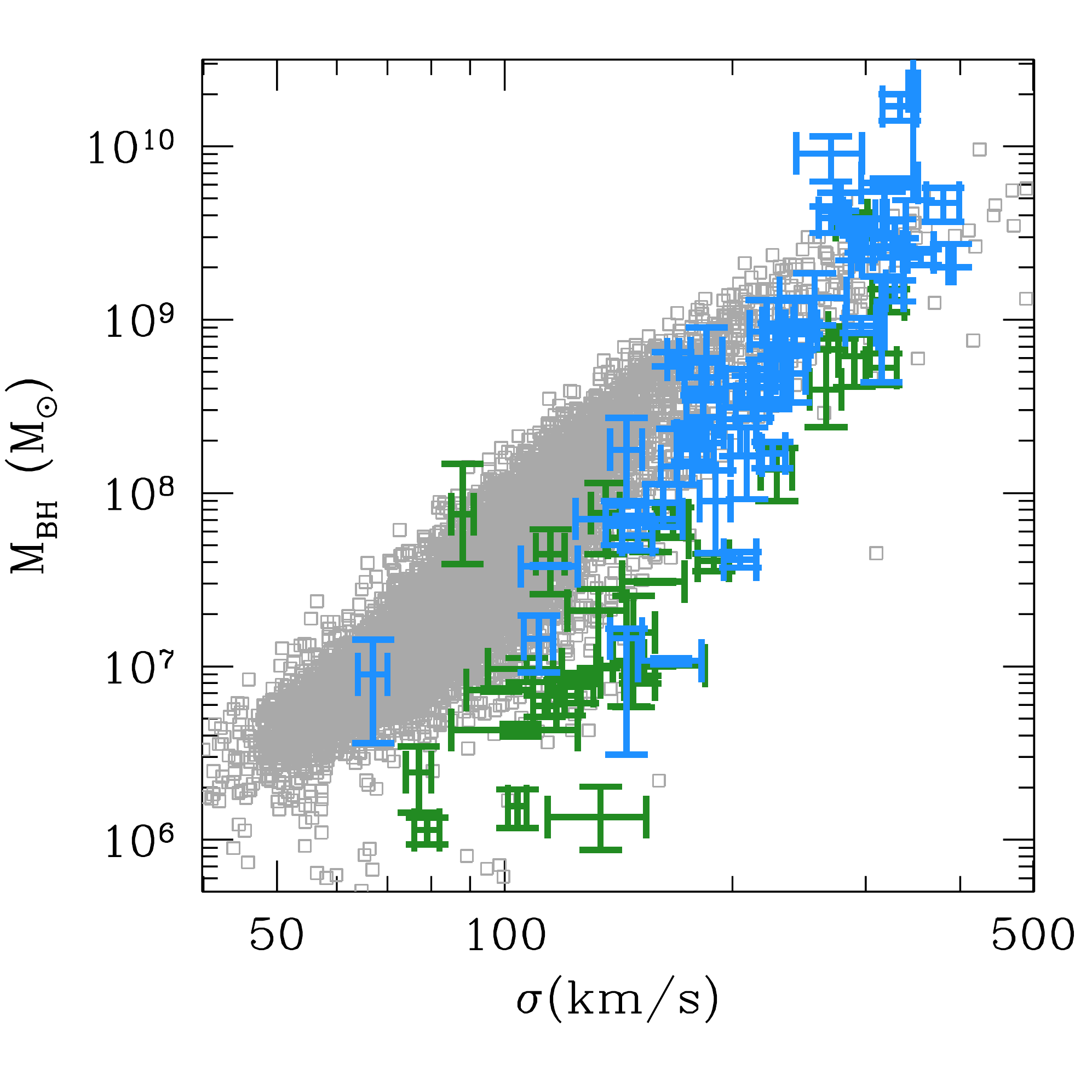}
\caption{BH mass versus bulge velocity dispersion at $z=0$ for non-interacting galaxies (grey).  Only 10000 random galaxies are shown. {We report also all BHs from Tables 2 and 3 in Kormendy \& Ho 2013, using blue symbols for the subset used by the Authors for establishing the BH mass versus bulge velocity dispersion relation. }Low-mass early-type galaxies in Horizon-AGN tend to have larger size than real ones, therefore the velocity dispersion is underestimated for these galaxies.}
\label{fig:BH_sigma}
\end{figure}

In Fig.~\ref{fig:BH_bulge} we turn to BH mass versus bulge mass. This is one of the ``classical" scalings that seem to have small scatter in the observations ($<0.3$ dex), at least for classical bulges, and which is widely extrapolated, also at high-redshift, and also when bulge/disc decompositions are not available, to study the co-evolution of BHs and galaxies. We start by highlighting that indeed the measurement of a univocal bulge mass is tricky. In our simulations we can define either a kinematical bulge mass, where bulge particles are those with a tangential velocity component (in cylindrical coordinates with the spin of the galaxy as the axis of symmetry) smaller than the non-tangential velocity component, or perform a bulge/disc decomposition, with either two Sersic profiles, one with $n=1$ for the disc, and one with $n=4$ for the bulge, or two Sersic profiles, both with $n=1$, or one with $n=1$, and one with $n=[1,4]$, keeping in mind that our physical resolution, 1~kpc, limits our ability to perform the decomposition in small or low-mass galaxies. 

The difficulties of defining a bulge mass is apparent in Fig.~\ref{fig:BH_bulge} and they are also discussed in Appendix A of \cite{2012MNRAS.420.2662D} for simulations and by \cite{2014ApJ...780...70L} for observations (see also Lasker et al. 2015).  We note that the bulge masses are taken from Kormendy \& Ho (2013), therefore, when the galaxy mass corresponds to the bulge mass, they are larger by 0.33 dex than the stellar masses shown in Fig.~\ref{fig:BH_gal}, because of the difference in zeropoint described by Reines \& Volonteri (2015). Despite the intrinsic difficulties, the match between Horizon-AGN BHs and the observational sample is good, where the comparison can be performed meaningfully (this was to be expected, as we reproduce correctly the BH mass function that is observationally determined through this correlation, as well as the stellar mass functions, see Kaviraj et al, in prep).  The logarithmic slopes for the various fits, including only objects with $M_{\rm bulge}>10^{9} \msun$, vary from 1.05 for the kinematical bulge mass, 0.75 for the two Sersic profiles with $n=1$ and $n=4$, 1.0 for both the two Sersic profiles with $n=1$, and $n=1$ and  $n=[1,4]$. 

In summary, the tighter BH-bulge mass relation is well reproduced in Horizon-AGN, while the broader BH-stellar mass scaling(s) are in worse agreement.   The light green triangles in Fig. 5 are almost entirely the same points that appear in Fig. 6. If we had restricted our analysis to this subsample, composed of galaxies with dynamically measured BH mass, which tend to be dense and bulge dominated, we would have a good match between simulation and observations also in Fig. 5, i.e. in BH-galaxy mass. In fact, our simulation and all other similar simulations have been tuned to match the  BH-stellar (bulge) mass correlation for the subsample of galaxies that works well (green triangles in Fig. 5, blue symbols in Fig. 6).  Such galaxies, however, are a biased subsample of the real universe \citep[see also][]{2016MNRAS.tmp..465S}. By tuning the simulations to reproduce a correlation that is tight only for a specific class of galaxies, there is the risk of failing to obtain the full view of the BH/galaxy population. One way to see this, is that relatively low-mass BHs in intermediate-mass galaxies that are not bulge-dominated, high-velocity dispersion galaxies, are not present in the simulation. 

Alternatively, we could see this as a problem with galaxy properties. Galaxies with stellar masses such that they should typically host $\sim 10^{7}\, \msun$ ÊBHs instead harbour BHs with mass ten times or more larger. At the same time the simulation reproduces, however, the observed BH mass versus bulge mass. The simulation may therefore over-predict bulge masses and  bulge-to-disc ratios.  We have compared our simulated galaxies to distributions of bulge-to-total ratios, \cite{2013MNRAS.432.1768K} for early-type galaxies and \cite{2009MNRAS.393.1531G} for disc galaxies, and \cite{2014MNRAS.441..599B} for a morphological mix. Simulated galaxies under-estimate the fraction of pure bulges, but over-estimate bulge-to-total ratios for galaxies with mass $10^{10} \lesssim M_{\rm gal} \lesssim 10^{11} \msun$. We stress, however, that this should be considered a qualitative comparison, as observational bulge-disc decompositions suffer from the same problems that we encountered in defining uniquely a bulge mass, plus issues related to inclination, surface brightness limits, spatial resolution and signal-to-noise ratio  \citep[see, e.g., the discussion in][]{2011ApJS..196...11S}.

Although merger histories and environmental effects induce some scatter in mock BH-galaxy relations, internal galactic  processes, such as gas turbulence, variation in the compactness, which are not captured because of limited resolution, may also influence the scatter.  The scatter given by different merger histories arising from cosmological initial conditions appears insufficient to explain the observational scatter, especially in  $M_{\rm BH}$ at fixed $M_{\rm gal}$, therefore these unresolved internal processes are likely to be an important contribution to the scatter. A similar behaviour  can be seen in other cosmological hydrodynamical simulations \citep[Fig. 4 and 10 in][respectively]{2015MNRAS.452..575S,2015MNRAS.446..521S}. 

We end by studying the BH mass versus stellar velocity dispersion, $\sigma$, in Fig.\ref{fig:BH_sigma}. This is also a benchmark scaling, and it is considered to be one of the best predictors for BH masses, i.e., to have a small intrinsic scatter ($<0.3$ dex).  This is perhaps the most problematic for Horizon-AGN. If we fit a single logarithmic slope to the BH mass versus stellar velocity dispersion, we find a value of 4.02 for galaxies with $\sigma>60 \kms$, compared to 4.38 found by Kormendy \& Ho on a culled subsample of the galaxies shown in Fig.\ref{fig:BH_sigma}.  The main reason is that velocity dispersions in the simulation are underestimated at $M_{\rm bulge}<10^{11} \msun$ ($\sigma<200 \kms$), and overestimated above, when comparing the observed Faber-Jackson relation \citep{Bernardi2003}. The dearth of BHs with $M_{\rm BH}<5\times 10^{7}\, \msun$ in intermediate mass galaxies highlighted in the discussion of Fig.~\ref{fig:BH_gal}  also contributes, but given the good agreement in the $M_{\rm BH}$ versus $M_{\rm bulge}$ relation, it is a minor contribution, for this specific comparison.  One contribution to the discrepancy is that galaxy sizes in the simulation are larger than for observed galaxies with mass $\lesssim10^{11} \msun$. In fact, the galaxy sizes are in good agreement with observed low-mass late-type galaxies, but are larger, by a factor of $\sim 3-4$, than observed low-mass early-type galaxies (Dubois et al., in prep.). Since the complete sample is a mix of late- and early-type galaxies, there appears to be an overall mismatch at $\sigma<200 \kms$ of a factor $\sim \sqrt 2$ in the velocity dispersion (roughly corresponding to an average overestimate of the galaxy size of a factor of 2).  The velocity dispersion is also sensitive to the dark matter halo central profile, which depends not only on dark matter halo mass but also on a subtle interplay with the baryons and AGN feedback. Indeed, the central dark matter densities in Horizon-AGN are lower than for the Horizon-noAGN simulation, which differs only in that AGN are not included in the latter. We defer a more detailed study quantifying the exact contribution of this effect as a function of galaxy mass
and assembly history to another paper (Peirani et al., in prep.).

\section{BHs in halos and subhalos: the effect of stripping}

Some BHs in local galaxies appear to have masses much larger than expected on the basis of the stellar or bulge mass (e.g., NGC~4486B, NGC~1277, M60-UCD1, NGC~4342, NGC~4291), and a possibility is that the host galaxy was originally more massive and lost some of its mass because of stripping \citep{2008MNRAS.384.1387V,2014Natur.513..398S}, although this is unlikely at least for some cases \citep[NGC~4342, NGC~4291;][]{2012ApJ...753..140B}.

\begin{figure}
\centering
\includegraphics[width=\columnwidth,angle=0]{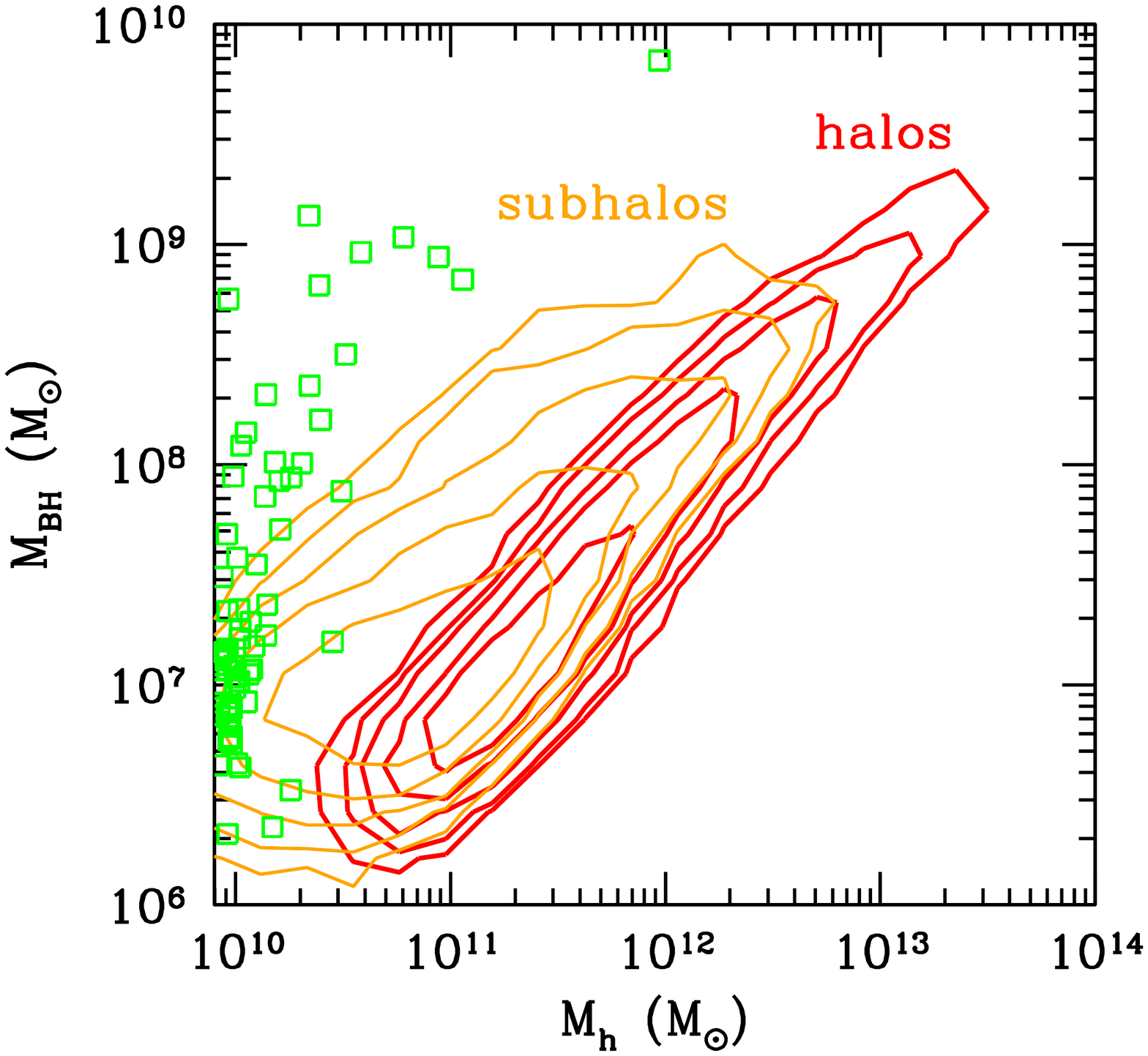}
\vspace{-2cm}
\includegraphics[width=\columnwidth,angle=0]{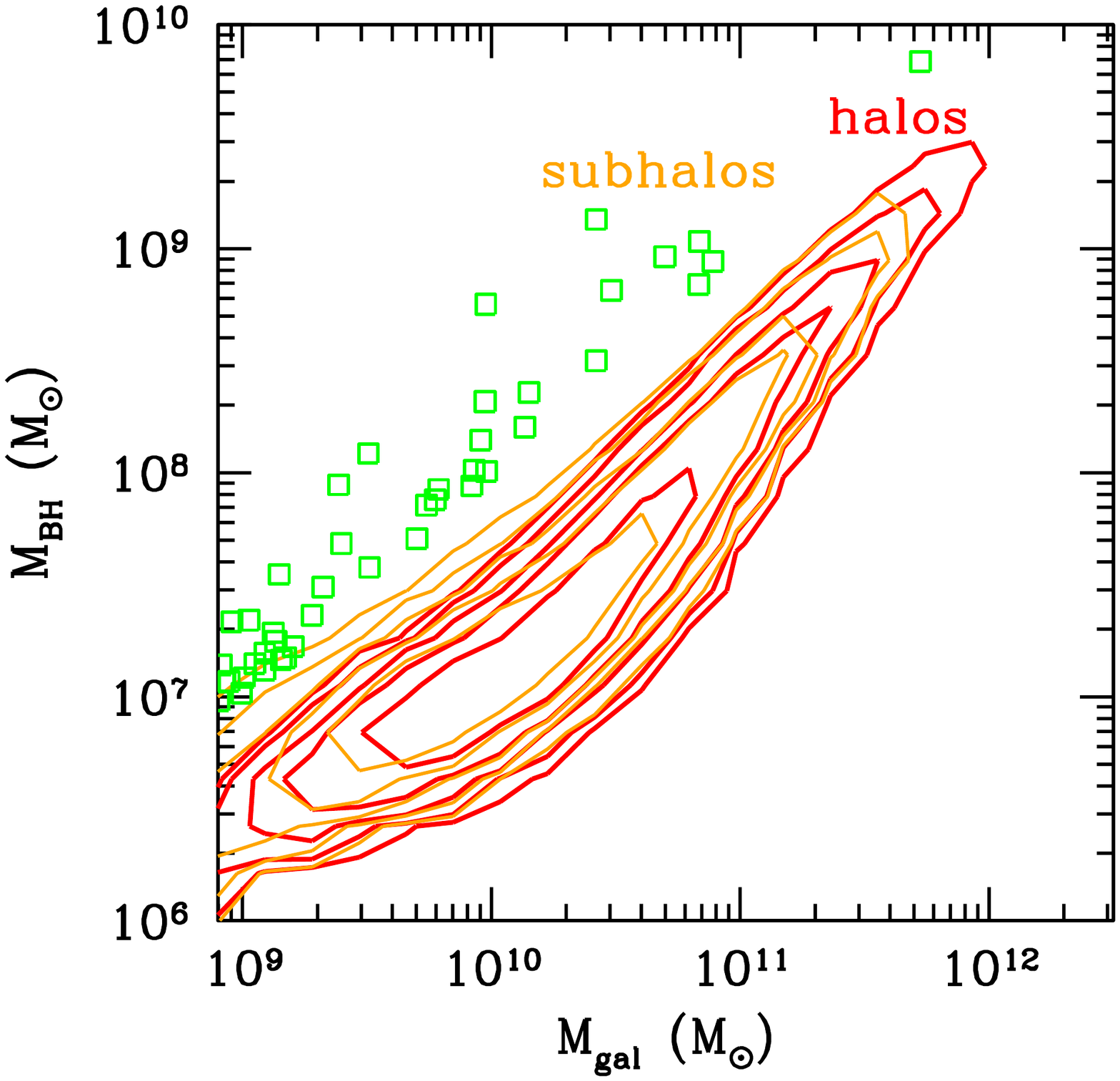}
\caption{BH mass versus halo mass (top) and versus galaxy mass (bottom) at $z=0$. Red: main halos; orange: sub-halos. Contours are based on ten equally-spaced logarithmic levels, dropping the five lowest levels for clarity.  At late cosmic times a population of BHs in stripped sub-halos emerges, where the BH is over-massive at a given halo mass. In reality, the halo is under-massive, having lost part of its mass. Stripping, however, is seldom effective all the way to the stellar distribution of galaxies. We highlight with green squares the few galaxies where the BH mass is $>0.01\times M_{\rm gal}$. All such galaxies reside in sub-halos.}
\label{fig:halo}
\end{figure}

We investigate here how stripping in sub-halos affects the connection with dark matter halos and stellar masses. In the top panel of Fig.~\ref{fig:halo} we show how sub-halos tend to host larger BHs (at fixed dark matter halo mass) compared to halos at $z=0$. This is an effect that, although present also earlier, builds up with cosmic time. Only very seldom, however, stripping is effective all the way into the stellar distribution and creates over-massive BHs (at fixed stellar mass). Normally the ratio between the BH and galaxy mass is between 0.002-0.005 for elliptical galaxies and bulges \citep{MarconiHunt2003,Haring2004,2013ARA&A..51..511K}, and can be 20-60 times lower for more active or more disky galaxies \citep{2010ApJ...721...26G,2015ApJ...813...82R}. We define a BH to be over-massive if its mass is $>0.01\times M_{\rm gal}$. 

The few subhalos with over-massive BHs are shown as green squares. Although we note that {\it there are no main halos hosting over-massive BHs} based on the definition given above, the cases where the BH is over-massive because of stripping are few.  Only 0.2 per cent of the entire galaxy population have a BH with mass $>0.01\times M_{\rm gal}$, and this fraction increases only by a factor of 3 if we restrict the sample to galaxies located within sub-halos.  At higher redshift, over-massive BHs are 0.10 per cent of the full population at $z=3$ and $z=2$, and 0.07 per cent at $z=1$. Restricting the sample to subhalos, the percentages change to 0.08, 0.11 and 0.17 at the three redshifts.  At early times most over-massive BHs are hosted in normal halos: BHs and galaxies are still growing -- and not always exactly at the same pace.

This is short by an order of magnitude compared with observations, where out of $\sim$ 135 galaxies with dynamical BH mass measurements, $\sim$ 8 galaxies have been suggested to be over-massive at $3-\sigma$ above the BH-bulge mass correlation  \citep[][but see Savorgnan \& Graham 2015]{2015ApJ...808...79F}. \nocite{2015arXiv151105654S}  \cite{2012ApJ...753..140B} proposed that over-massive BHs where the host has not been stripped can be explained by non-standard paths of BH-galaxy co-evolution \citep[see also][]{2013MNRAS.431L..38F,2014ApJ...780L..20T,2015ApJ...808...79F}. Some of the proposed paths include a sequence of accretion events dominated by low angular momentum gas, boosting BH growth, and causing star formation to occur in a compact configuration. Once the BH has grown sufficiently, AGN feedback prevented additional BH and galaxy growth. In a similar vein, these galaxies and their BHs simply do not experience any growth by mergers after $z\sim3$, remaining structurally untouched and without further mass and size increase, but recording a time when, statistically, the ratio between BH and bulge mass was larger than today.  Keeping in mind that scatter in the simulated BH-galaxy relations is less than in observations, we have looked for BHs that are ``almost" over-massive, $>0.004\times M_{\rm gal}$, and indeed found some cases where the BH is not hosted in a sub-halo, and the galaxy growth has nearly stopped at $z\sim 2-3$.  These cases remain very few, at 10 per cent of the ``almost" over-massive BHs, which in turn represent the 2.5 per cent of the full population. 

In summary, a very small fraction of BHs appear over-massive with respect to the stellar mass of their host because the latter has been stripped, being a sub-halo of a larger halo. However, {\it all} the BHs with $M_{\rm BH}>0.01\times M_{\rm gal}$ we identified in our simulation at $z=0$ are in sub-halos \citep[see also][]{2016arXiv160204819B}. If we are more generous in our definition of over-massive, $>0.004\times M_{\rm gal}$, then a larger population emerges, including also some cases where tidal stripping is not the culprit.  To put in context the fraction of over-massive BHs in the simulation and in observations, it is also important to remark that many (if not most) galaxies for which Êdynamical BHÊmass measurements exist are located in galaxy clusters where tidal stripping and galaxy-galaxy interactions Êare expected to beÊmore important (especially in the centre of galaxy clusters) than in a simulation spanning a wide range of environments such as the one presented here.

\begin{figure}
\centering
\includegraphics[width=\columnwidth,angle=0]{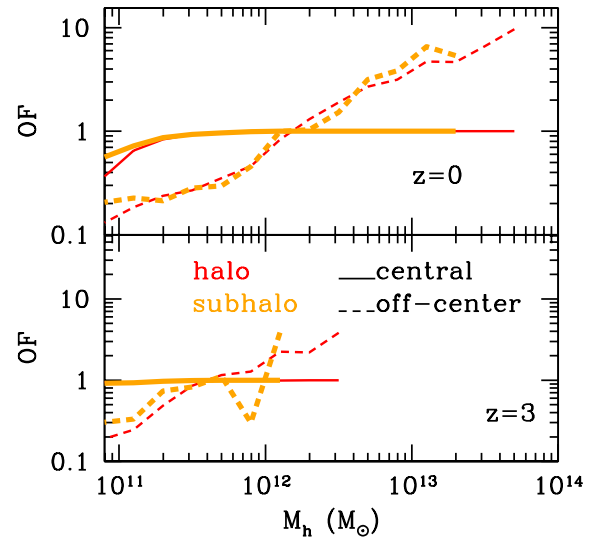}
\caption{Occupation fraction of BHs as a function of halo mass at $z=0$ (top) and $z=3$ (bottom). We separate main halos (red, thin curves) and sub-halos (orange, thick curves). Solid lines represent central BHs, i.e., BHs located within 10 percent of the virial radius. The occupation fraction of central BHs reaches unity at halo masses $\sim 10^{10} \msun$ at $z=3$ and  $\sim 10^{11} \msun$ at $z=0$, thus tracing the mass growth of halos. The occupation fraction of subhalos, at a given mass, is higher than that of halos because of stripping: the initial halo mass was larger than at the present time, therefore the OF traces the initial halo properties (cf. Fig.~\ref{fig:halo}). Dashed lines represent off-centre BHs which are located within the virial radius of a given halo and are not associated with any sub-halo. Tens of off-centre BHs can be found in the most massive halos. }
\label{fig:halo_OF}
\end{figure}

\begin{figure}
\centering
\includegraphics[width=\columnwidth,angle=0]{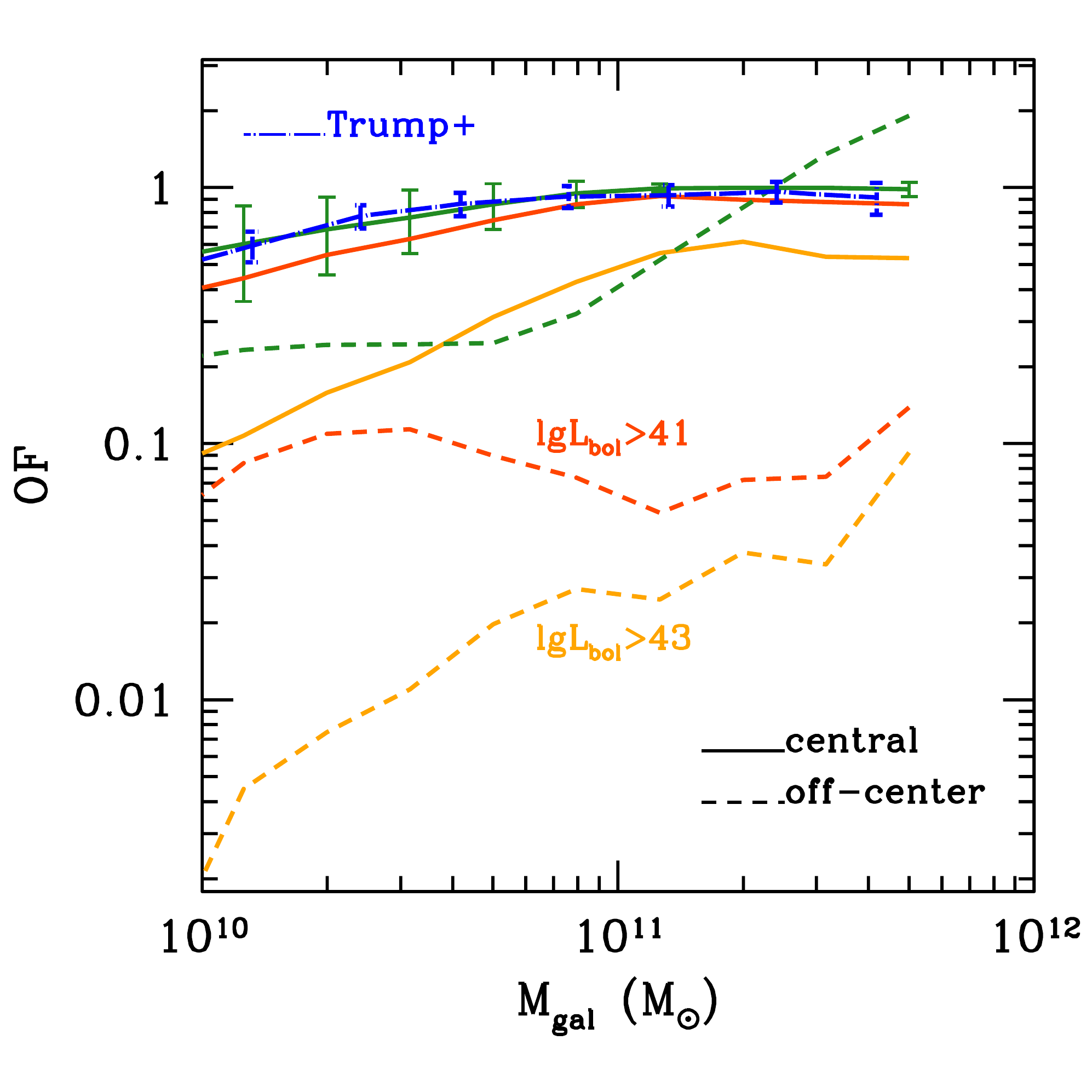}
\caption{Occupation fraction of BHs as a function of galaxy mass at $z=0$ (dark green). Solid lines represent central BHs, i.e., the most massive BH located within 10 percent of the virial radius and $2\times$ the galaxy effective radius. The blue dot-dashed curve is the observational estimate by \citet{2015ApJ...811...26T}. Dashed lines represent off-centre BHs which are not associated with any satellite galaxy. In principle the number of off-centre BHs can be larger than unity in a given galaxy (e.g., as a result of a multiple mergers). The red and orange curves show the occupation fraction for BHs above $10^{41}\, {\rm erg\, s^{-1}}$ and $10^{43} \,{\rm erg\, s^{-1}}$ respectively, to highlight the difficulty of finding BHs in low-mass galaxies.  If we included a correction for suppressed radiative efficiency when the Eddington ratio is below 0.01, the occupation fractions above the two luminosities would be 60-80 per cent of those shown here. Obscuration can also affect up to 80 per cent of faint AGN. The off-centre BHs with  $L_{\rm bol}>10^{43} \,{\rm erg\, s^{-1}}$ are normally located $<10$ kpc. Those stranded in the halo diffuse hot gas have $L_{\rm bol}<10^{39} \,{\rm erg\, s^{-1}}$.  Errorbars are shown only on one of the curves for central BHs for clarity. In the case of off-centre BHs, the scatter around the mean is about 0.3-1 dex.}
\label{fig:gal_OF}
\end{figure}

\begin{figure}
\centering
\includegraphics[width=\columnwidth,angle=0]{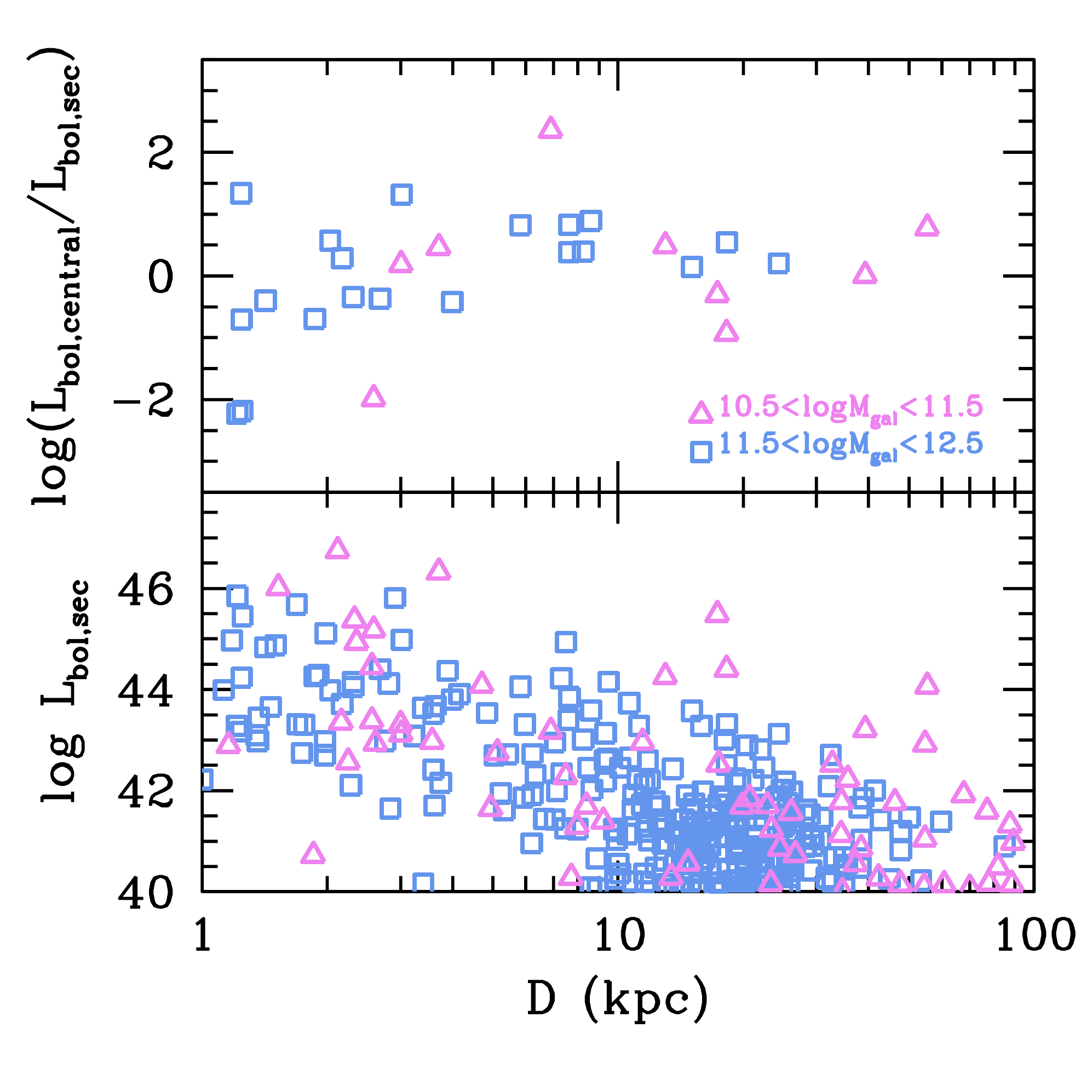}
\caption{Top: ratio between the luminosity of the central AGN and the secondary AGN, when both AGN have $L_{\rm bol}>10^{43} \,{\rm erg\, s^{-1}}$. Bottom:  luminosity of the secondary AGN as a function of distance from the galaxy centre.  The incidence of luminous dual AGN increases as the separation decreases, with a tail of faint BHs at large distances from the centre (the typical Eddington rate is $<0.01$ for BHs at distance $>10$ kpc). The former have recently taken part in a merger. The latter are BHs wandering in the halo diffuse hot gas, with negligible accretion.}
\label{fig:dual}
\end{figure}

\begin{table*}
\caption{Fraction of galaxies above a given mass with off-centre or dual BHs or AGN. The second and third column give the fraction of galaxies hosting an off-centre BH, or AGN above $L_{\rm bol}>10^{43}\,{\rm erg\, s^{-1}}$. The fourth and fifth column give the fraction of off-centre BHs or AGN that are in dual systems, i.e. a central BH or AGN, matching the same luminosity criterion, exists. The fourth and fifth column give the fraction of dual BHs or AGN over the total number of galaxies within the mass bin.}
\begin{center}
\begin{tabular}{|c|c|c|c|c|c|c|c}
$M_{\rm gal} (\msun)$ & off-centre  & off-centre  & dual BH  &  dual AGN & dual BH  &  dual AGN (total) \\
($10^{11}\msun$) &  BH  & AGN   & (off-centre)  &(off-centre)     & (total )  & (total) \\
\hline
$>0.1 (z=0)$ & 0.36 & 0.01 & 0.38  & 0.13 & 0.14  & 0.001\\
$>1 (z=0)   $ & 1.0   & 0.04 & 1.0    & 0.27 & 1.0 & 0.009\\
\hline
$>0.1 (z=2)$ & 0.60 & 0.15 & 0.56  & 0.17 & 0.35  & 0.02\\
$>1 (z=2)  $  & 1.0   & 0.21 & 0.82  & 0.53 & 1.0    & 0.11\\

\end{tabular}
\end{center}
\label{table:dual}
\end{table*}

\section{The occupation fraction of central and off-centre BHs}

In massive galaxies where a BH has been looked for, one was found.  However, the situation differs for low-mass galaxies.  Faint AGN are now being found in many low-mass galaxies \citep{2013ApJ...775..116R,2015ApJ...799...98M,2015ApJ...805...12L,2015ApJ...811...26T,2015arXiv150703170P,2016ApJ...817...20M}, but there are only upper limits for BHs in NGC~205 \citep{Valluri2005}, M33 \citep{Gebhardt2001}, Fornax \citep{2012ApJ...746...89J}. The fraction of halos or galaxies hosting BHs as a function of galaxy mass and properties (BH occupation fraction, OF) is an important diagnostic to learn about the first BHs, in a way similar to near-field cosmology and the first galaxies \citep{VLN2008,2010MNRAS.408.1139V}.  The fraction of galaxies hosting AGN as a function of galaxy mass and properties is a lower-limit to the BH OF, and, additionally provides information on AGN fueling. In Horizon-AGN BH seeding has been performed in a relatively simple way, not directly based on models of BH formation in the early Universe, and it has been stopped at $z=1.5$.  The choice of halting BH formation at this redshift does not affect our conclusions, as all well-resolved galaxies at the end of the simulations have at least one progenitor at $z=1.5$. Although we cannot derive constraints on BH formation scenarios, we can test qualitatively the trends with galaxy mass, i.e., the expectation that low-mass galaxies have more trouble forming and keeping BHs, and feeding them. 

The OF of BHs as a function of halo mass is shown at two different redshifts ($z=0$ and $z=3$) in Fig.~\ref{fig:halo_OF}. All halos with mass $>10^{11} \msun$ at $z=0$ and $>3\times10^{10} \msun$ at $z=3$ host a central BH. The OF is higher at low halo mass for sub-halos, for reasons similar to those discussed in the previous section. Before the halo became a sustructure, it was more massive, and this is the information recorded by the OF.  

There are also halos that host off-centre BHs. To find off-centre BHs, we first assign all the central BHs, within $0.1\times R_{\rm vir}$, and remove them from the list. We then consider the BHs which do not belong as centrals to any halo, and assign an off-centre BH to a halo if it is within $R_{\rm vir}$. We start from sub-structures and proceed then to main halos.  These off-centre BHs are generated by mergers during the hierarchical assembly of halos. Some of them are the result of a recent merger, and they are on their way to joining and coalescing with the central BH, and may have very high luminosity, through merger-driven gas inflows. In halos with masses $3\times 10^{12}<M_h<3\times 10^{13} \msun$ these merger-driven off-centre AGN are located within 10 kpc from the halo centre, and have luminosities $\sim 10^{43}-10^{47}$ erg/s.  Other off-centre BHs are stranded in the outer parts of  halos and have very low luminosities. In halos with masses $3\times 10^{12}<M_h<3\times 10^{13} \msun$ most off-centre BHs are located at $\sim 100$~kpc from the halo centre and have luminosities of $\sim 10^{36}\, {\rm erg\, s^{-1}}$. The luminosities are low because these BHs are surrounded by hot, low density gas. By $z=0$ the most massive halos can host tens of these BHs, as shown with dashed lines in Fig.~\ref{fig:halo_OF}.

Fig.~\ref{fig:gal_OF} estimates the OF of central and off-centre BHs at $z=0$ as a function of galaxy stellar mass, more easily measurable in observations. We compare the OF of central BHs in Horizon-AGN to that estimated by \citet{2015ApJ...811...26T}, finding a very good agreement. In this figure we also show how the OF changes when we assume that only sufficiently luminous AGN can be used as a signpost for detection. We impose two different thresholds in bolometric luminosity,  $10^{41} \,{\rm erg\, s^{-1}}$ and $10^{43} \,{\rm erg\, s^{-1}}$. With a threshold of $10^{43} \,{\rm erg\, s^{-1}}$, close to the value typically used for defining an AGN, the detectable OF of central BHs decreases by a factor of $\sim 2$ in massive galaxies, but below $10^{10} \msun$, the AGN OF is about 1 per cent of the BH OF.  The fraction of off-centre AGN is similarly suppressed by a factor of $\sim$ ten, so that the apparent off-centre AGN occupation fraction is always below 10 per cent. We have used as radiative efficiency 10 per cent, to be consistent with the {\sc ramses} implementation, but, including in post-processing a correction for suppressed radiative efficiency when the Eddington ratio is below 0.01 \citep{Merloni08}, the occupation fractions above the two luminosities would be 60-80 per cent of those shown here. We have also not included a correction for obscuration, which affects up to 80 per cent of AGN at faint luminosities \citep[e.g.,][]{Lafranca2005,2014ApJ...786..104U}.

Finally, a fraction of the off-centre BHs are dual systems, i.e., the galaxy hosts both a central BH and an off-centre BH, where we define the central BH as the most massive between the two, when they are at comparable distance. In some cases both BHs are active as AGN. Figure~\ref{fig:dual} shows the luminosity of the central and off-centre AGN, as well as the ratio of the two luminosities for the cases where $L_{\rm bol}>10^{43}\,{\rm erg\, s^{-1}}$ for both AGN, at $z=0$. The ratio between the two luminosities is within $\sim$ one decade. This is because of two reasons. Firstly, we impose a minimum luminosity cut to both. Secondly, only mergers with a mass ratio sufficiently close to unity trigger AGN activity on both BHs \citep[e.g., $>1:6$ in][]{2015MNRAS.447.2123C}. Therefore, as the mass ratio of the merging galaxies is not dissimilar, and the BH mass scales with the galaxy mass, the luminosities of the two BHs, during the active phase, will also not be too dissimilar. 

The general trend is that more luminous duals tend to be at closer distances from the centre \citep[see also][]{VW2012,2015arXiv151008465S}. In the bottom panel a population of low-luminosity, large-distance duals emerges (here we define as ``secondary" the non-central AGN). Once again, these are BHs stranded in galaxy outskirts, where gas is hot and thin. In Table~\ref{table:dual} we report the fractions of off-centre BH and AGN (dual or single), the fraction of off-centre BH and AGN that are duals, and the fraction of duals over the total number of galaxies above two galaxy mass thresholds ($10^{10}\msun$ and $10^{11}\msun$) at $z=0$ and $z=2$. The fraction of off-centre, but especially of dual AGN, increases at higher redshift, as suggested by \cite{2014ApJ...789..112C}. We stress, however, that this fraction is based only on meeting a luminosity criterion. Observational searches, instead, are based on selecting for a velocity shift caused by Doppler motion in the AGN and galaxy spectra \citep{2009ApJ...698..956C}, therefore our results should be only taken as an indication of astrophysical trends rather than a prediction directly comparable to observations. We refer the reader to discussions in \cite{2014ApJ...789..112C}, \cite{2015ApJ...813..103M}, \cite{VW2012}, \cite{2013MNRAS.429.2594B}, \cite{2015arXiv151008465S}, and references therein for caveats that enter into making detailed comparisons between simulations and observations.  We also note that, although we have tried to limit our analysis only to massive galaxies that are in better agreement with observations, based on the discussion in section~3, simulations reproduce only partially a realistic feeding cycle for BHs, therefore these should be considered as qualitative trends.

\begin{figure}
\centering
\includegraphics[width=\columnwidth,angle=0]{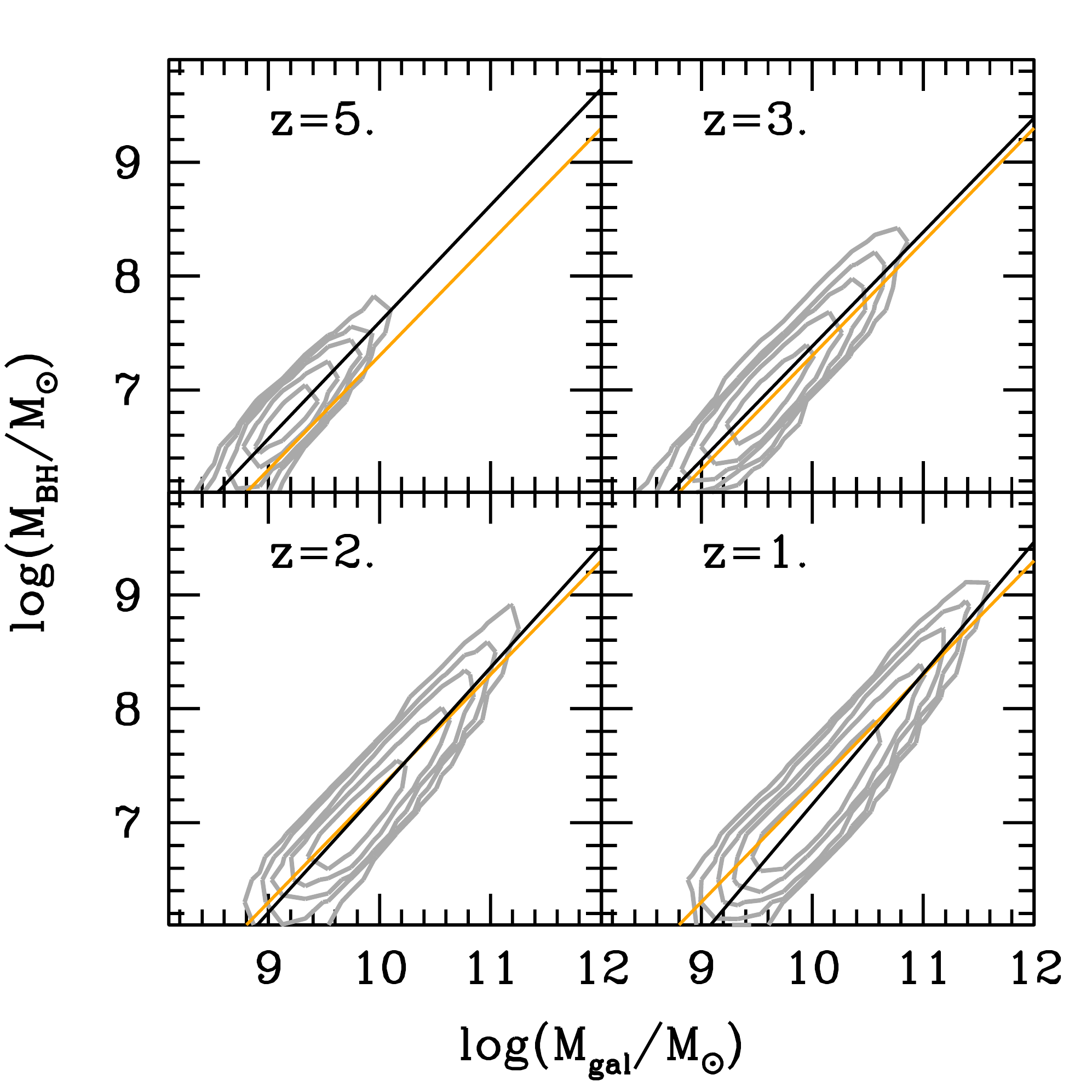}
\caption{BH mass versus total galaxy mass at different redshifts for galaxies hosted in halos with mass $>8\times 10^{10} \msun$.  The orange line, meant to guide the eye, represent the locus where the BH mass is two thousandths of the galaxy mass. The black line is the best fit linear correlation.}
\label{fig:gal_z}
\end{figure}

\section{Growing BHs in galaxies}

In the following we dig deeper in how BHs change over time as a function of BH and host galaxy and halo masses. We also identify the dominant population in terms of mass growth and accretion properties, and its evolution over cosmic time. 

\subsection{Correlations between BHs and galaxies at $z>0$}

As hinted in section 3.2, the correlations between BH mass and galaxy properties as a function of redshift are not well constrained observationally, nor robustly predicted theoretically. Observations are fraught with difficulties, both in measuring BH masses and galaxy properties. The BH masses are estimated through indirect methods, radically different from the dynamical techniques used at $z=0$, and with larger systematic uncertainties \citep[e.g.,][]{vestergaardpeterson2006,Shen2013}.  The bulge mass is very seldom measurable, even on HST data \citep{2013ApJ...767...13S,2015ApJ...799..164P}. The total stellar mass is the property more readily measurable, but only for type-2 AGN where the host can be detected \citep[e.g.,][]{Merloni2010,2014MNRAS.443.2077B,2015ApJ...802...14S}. Additionally, selection biases \citep{Lauer2007b} tend to cause flux-limited samples to select BHs that lie above the relations between BH and galaxies. In general, observed samples suggest that the ratio between BH and galaxy mass is higher at higher redshift, but once the selection biases are taken into account this is not obvious anymore \citep{2011A&A...535A..87S,2014MNRAS.438.3422S}. 

Theoretically, different results have been obtained \citep{Robertson2005,2015MNRAS.454..913D,2015MNRAS.452..575S}, with either rigid scaling of the normalization, or with changes in the slope. We report in Fig.~\ref{fig:gal_z} the relation between BH and galaxy stellar mass at different redshifts in Horizon-AGN, including only BHs and galaxies hosted in halos with mass $>8\times 10^{10} \msun$.  The slope of the logarithmic relation varies as 1.02-1.00-1.08-1.15-1.08 from $z=5$ to $z=0$. The normalization decreases when the slope decreases and vice versa, so that, in the relevant mass range, there is very little evolution with redshift, and the relation hovers around a simple scaling such that $M_{\rm BH} \simeq 2\times 10^{-3} M_{\rm gal}$. 
 
 We also find very little scatter in the BH-galaxy mass relation, which, as discussed in section 3.4, is tighter than in observations at $z=0$. 

\begin{figure}
\centering
\includegraphics[width=\columnwidth,angle=0]{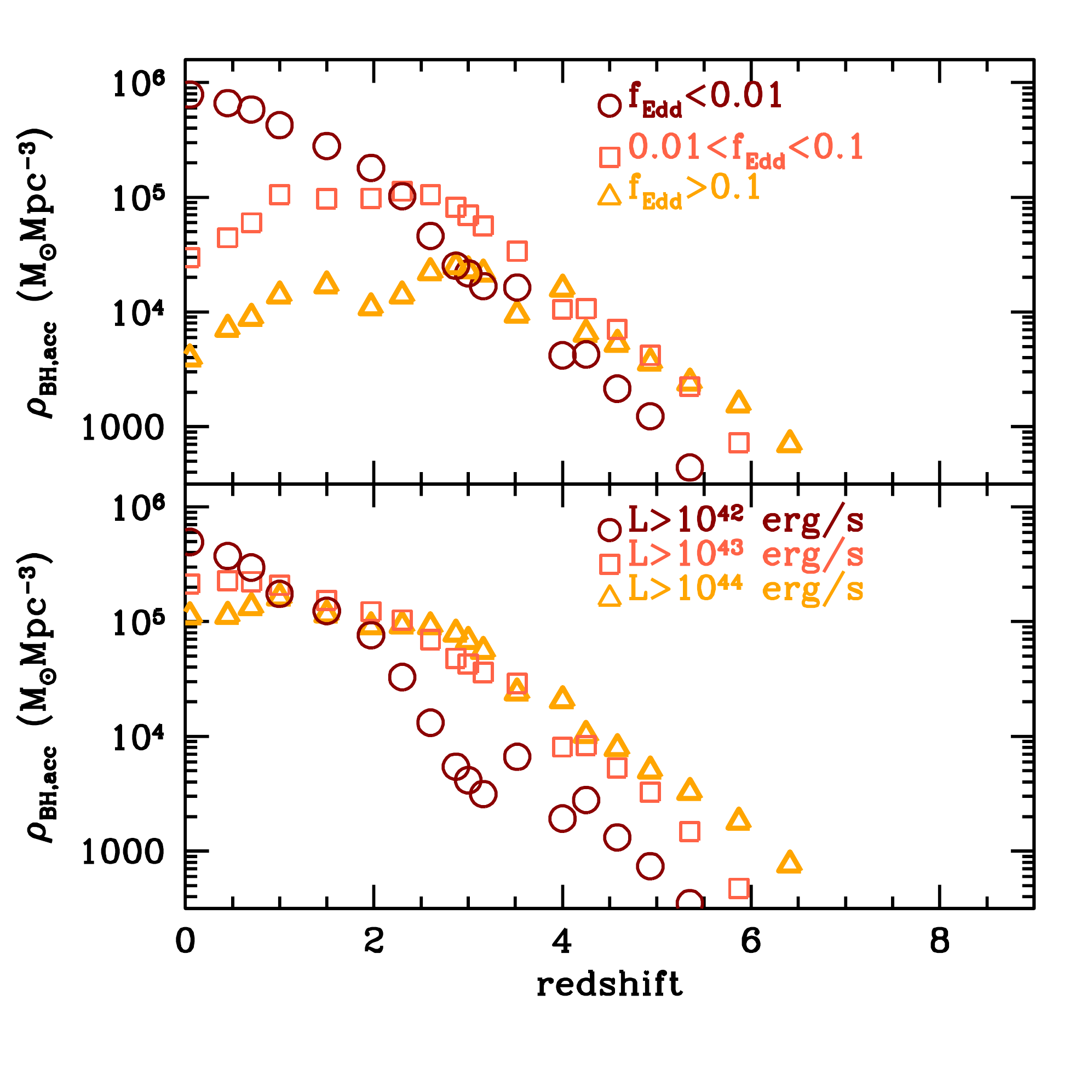}
\caption{Mass density in accreting BHs of different Eddington ratios (top) or luminosities (bottom). Until $z\sim 3$ a large fraction of BHs accrete mass in bright phases with high Eddington ratios, but at later times the mass growth is dominated by faint, low-Eddington rate accretors.}
\label{fig:rholum}
\end{figure}

\subsection{The transition between fast and slow accretors}
In section 3.2 we used the BH mass density as a constraint for the simulation. We can also use it as a tool to assess which BHs are growing most effectively and contribute most to the general population at different cosmic times. We show in the top panel of Fig.~\ref{fig:rholum} the contribution of BHs accreting above different Eddington ratios to the accreted BH mass density.  We recall that this is the mass density in accreting BHs at a given redshift, rather than the cumulative contribution integrated over redshift. Until $z=4$ almost all the BHs are fast accretors ($f_{\rm Edd}>0.1$), with about 50 per cent of the mass density locked in BHs accreting close to the Eddington rate. This is the population that dominates the luminosity function of quasars \citep{2006ApJ...648..128K}. Slow accretors take over afterwards, and the mass budget is dominated by BHs with $0.01<f_{\rm Edd}<0.1$. The slowest accretors ($f_{\rm Edd}<0.01$) prevail at $z<2$.  The luminosity density, combination of BH mass and Eddington ratio, however, is always roughly equally contributed by fast and slow accretors, as BH masses can only increase with time. 

A more observation-friendly way of seeing the contribution of different sources to the BH mass density is shown in the bottom panel of Fig.~\ref{fig:rholum}, where we bin in luminosity. Since, as noted above, the luminosity depends on both the Eddington ratio and the BH mass, an increasing function of time, the trends with redshift are similar, but not the same with respect to the Eddington ratio case. A high mass BH with a low accretion rate (slow accretor) can have the same luminosity as a low mass BH with a high accretion rate (fast accretor): binning in luminosity suffers from this degeneracy and limits its discriminating power. 

If we were to divide the contribution by BH mass, BHs with mass $<10^7 \msun$ dominate at $z>6$, BHs with mass $10^7<M_{\rm BH}<10^8 \msun$ dominate between $z=6$ and $z=2$ and BHs with mass $>10^8 \msun$ take over afterwards. We recall that the volume of Horizon-AGN, although very large, $2.86 \times 10^6 {\rm Mpc}^3$, is not sufficient to capture the rarest density peaks where the most luminous quasars at high-redshift are expected to reside \citep{2012ApJ...745L..29D,Dubois2013,2014MNRAS.439.2146C,2015arXiv150406619F}. BHs with masses $>10^8 \msun$ start to appear in substantial number only at $z<5$, but by $z=3$ they have already taken over in the total accreted mass budget, in agreement with the expected ``downsizing" scenario. As discussed above, downsizing is more clearly evident in the connotation of strong versus weak growth: the predominance of fast accretors at high redshift, and the transition to slow and weaker accreting BHs at low redshift.  

\begin{figure}
\centering
\includegraphics[width=\columnwidth,angle=0]{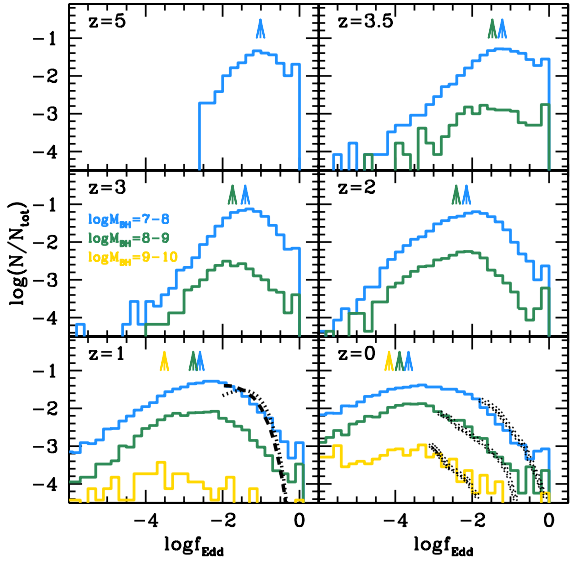}
\caption{Distribution of Eddington ratios for BHs of different masses. Arrows mark the mean value for each curve. The black dotted curve in the $z=1$ panel reports the distribution function  proposed by \citet{2015MNRAS.447.2085S} for $\log(M_{\rm BH}/M_\odot)=7.5$, with the dashed curve showing the obscuration-corrected version.   The black dotted curves in the  $z=0$ panel show the results found for low redshift AGN in the SDSS by  Heckman et al. 2004. The most massive BHs have the lowest Eddington ratio. At a given BH mass, the mean value (shown with an arrow) moves to lower Eddington ratios as redshift decreases.}
\label{fig:fedd}
\end{figure}

\subsection{Eddington ratios, specific accretion rates and duty cycles}
The accreted BH mass density is an integral quantity, and it is dominated by BHs at the knee of the luminosity function.  The progressive weakening of BH accretion, in a global sense, can be appreciated by looking at the distribution function of Eddington ratios for BHs of different masses at different redshifts (Fig.~\ref{fig:fedd}). The fraction of BHs with low Eddington ratios increases with cosmic time, and at a given BH mass the peak of activity shifts towards lower Eddington ratios with time. Note that the distribution is normalized in each bin to the total number of BHs at each redshift, thus giving an indication of the BHs comprising the bulk of the population.   We compare the distribution in Horizon-AGN with the analytical fit by \cite{2015MNRAS.447.2085S} obtained on samples  of type-1 AGN from three optical surveys (VVDS, zCOSMOS and SDSS) at $1<z<2$, for which the BH masses have been estimated from broad lines in single epoch spectra. They perform a maximum likelihood fitting of the bivariate distribution of Eddington ratios and BH mass,  allowing for a BH mass dependence in the  Eddington ratio distribution, restricted to $f_{\rm Edd}>0.01$. We show as a dotted curve their best model, which adopts a Schechter function for the Eddington rate distribution, and as a dashed curve their distribution corrected for obscuration, with the luminosity-dependent correction for absorption proposed by \cite{2014MNRAS.437.3550M}, both rescaled so that the total number of objects with $f_{\rm Edd}>0.01$ is the same as in the Horizon-AGN distribution. The agreement with the absorption-corrected distribution is very good up to $f_{\rm Edd}\sim0.5$, but the simulation predicts a larger population of fast accretors, with the small peak at $f_{\rm Edd}=1$ caused by the Eddington cap in the Bondi formalism.  At $z=0$ we compare the simulation to the distribution found by \cite{2004ApJ...613..109H} for low-redshift AGN in the SDSS. Their results for $\log(M_{\rm BH}/M_\odot)=7.5, 8.5, 9$, shown as polka-dot curves, match very nicely the simulation results, except at the highest $f_{\rm Edd}$ where, however, the statistics are low.

A second comparison we perform is with the distribution of specific BH accretion rate, i.e., BH accretion rate relative to the stellar mass of the host galaxy, as defined by \citet{Aird2012} and \citet{Bongiorno2012}: 
\begin{equation}
{\rm specific \,BHAR}=\frac{L_{\rm bol}}{10^{38} \, {\rm erg \,s^{-1}}} \left(\frac{0.002 M_{\rm gal}}{\msun}\right)^{-1}.
\end{equation}

\begin{figure}
\centering
\includegraphics[width=\columnwidth,angle=0]{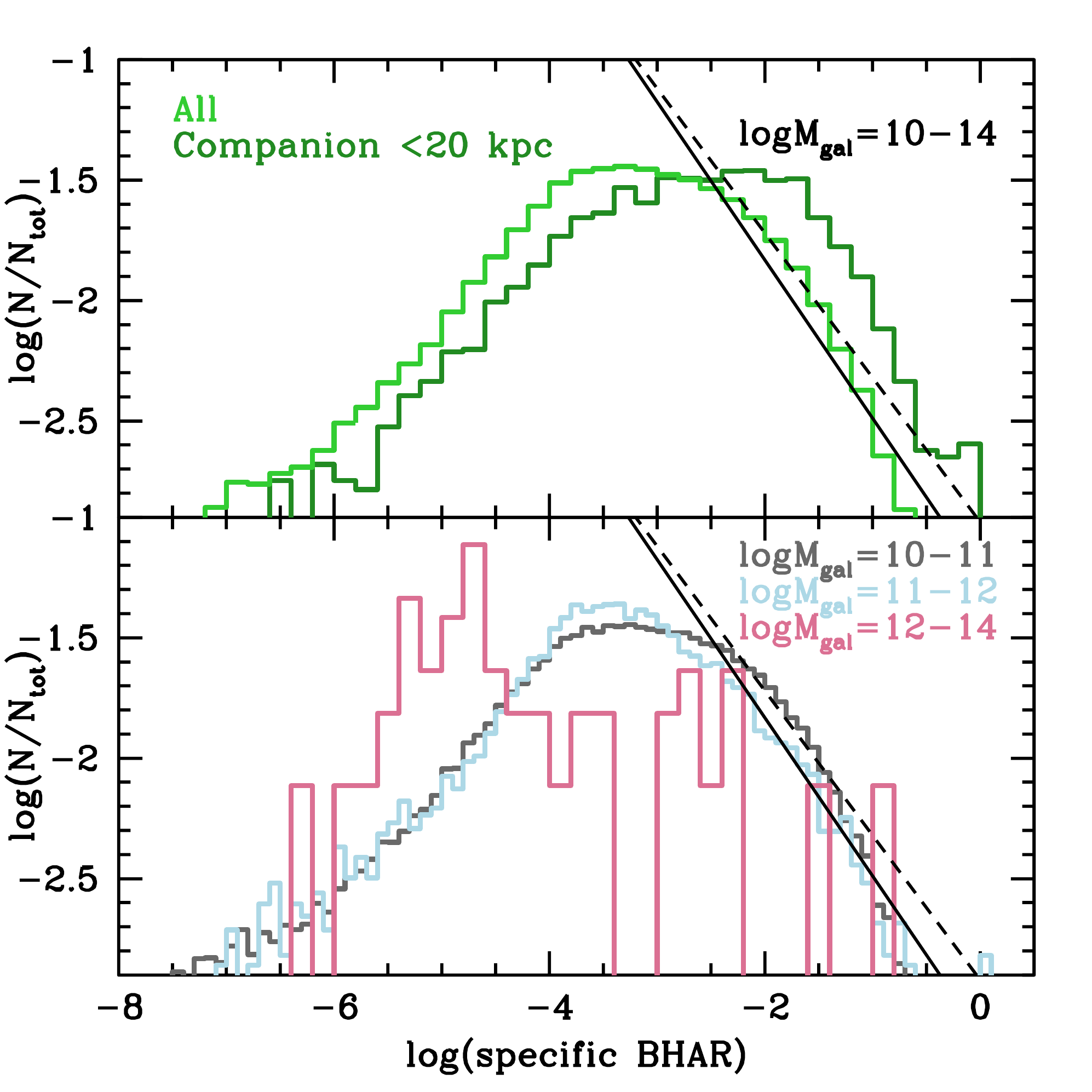}
\caption{Distribution of specific BH accretion rate, i.e., BH accretion rate relative to the stellar mass of the host galaxy, for BHs in Horizon-AGN for different cuts in galaxy mass. The specific accretion rate is a proxy for the Eddington ratio if BH mass scales with galaxy mass. With this more general definition, however, the comparison with observations is straightforwards, as one does not need to measure a BH mass (needed instead for the Eddington ratio). As a reference we show the distributions proposed by Aird et al. 2012 (solid line) and Bongiorno et al. 2012 (dashed line), found to be independent of host stellar mass. We also find that the specific BH accretion rate is nearly independent of galaxy mass (bottom panel), and we identify a specific BH accretion rate ($\log(f_{\rm Edd})=-3.4$) below which the power-law distribution breaks. The top panel shows that BHs in galaxies with close companions tend to have larger specific BH accretion rates (see also Volonteri et al. 2015).}
\label{fig:aird}
\end{figure}

This quantity is analogous to the Eddington ratio, if BH mass scales with galaxy mass, more easily obtained in observations, and it is less sensitive to uncertainties in estimates of the BH mass. Since in Horizon-AGN the BH mass scales tightly with the galaxy mass, the two quantities are indeed very similar, as we checked by using $f_{\rm Edd}$ instead of the specific BHAR. The specific BHAR distribution is shown in Fig.~\ref{fig:aird} for BHs in galaxies at $z=0-1$. In this case we normalise each distribution to the number of objects in each mass bin, to compare with  \cite{Aird2012}. The bottom panel of Fig.~\ref{fig:aird} shows that, in agreement with observations, there is not much dependence on the galaxy mass, once each mass bin is self-normalized. However, in this case we find a turnover, not found by  \cite{Aird2012}.   Therefore, the simulation predicts, at low Eddington ratios, a behaviour in between the power-law proposed by Aird et al. (2012) and Bongiorno et al. (2012), and the Schechter function proposed by \cite{2015MNRAS.447.2085S}: we do predict a turnover, but at lower values than found by \cite{2015MNRAS.447.2085S} and \cite{2012MNRAS.425..623L}, keeping in mind that the latter two papers have estimated $f_{\rm Edd}$, rather than the specific BHAR. Also, our estimate is for the whole BH population: we are not limited to type-1 AGN, nor to AGN above a luminosity or flux limit. Observations, instead, are necessarily impacted by selection effects, and this may explain the discrepancy in observational results. 

Since the mass dependence is negligible, in the top panel we group all galaxy masses, but we split the sample between isolated galaxies and galaxies with a companion within 20~kpc. For the full sample, except for a tail at $f_{\rm Edd}<10^{-5}$, the distribution is well approximated by a lognormal with $\mu=-3.2$ and $\sigma=1.02$. Separately, at $z=0$, $(\mu,\sigma)=(-3.42,1.02)$ and at $z=1$, $(\mu,\sigma)=(-2.44,0.81)$. For galaxies with a companion, the distribution shifts towards slightly larger Eddington ratios, $(\mu,\sigma)=(-2.54,1.17)$ at $z=0-1$, as found also for simulations of isolated galaxy mergers \citep{2015MNRAS.449.1470V}. Observations also find an enhancement of AGN activity in galaxies with companions \citep{2008AJ....135.1877E,2011ApJ...743....2S,2011MNRAS.418.2043E,2014AJ....148..137L,2012ApJ...746L..22K}.  A detailed discussion of the role of mergers in shaping the BH mass budget will be forthcoming in a companion paper (Dubois et al. in prep.), and compared to the result for star formation. For galaxies,  \cite{2015MNRAS.452.2845K} show that in Horizon-AGN the star formation enhancement during major mergers, with a mass ratio $>1:4$, is between 20-40 per cent, and that about a quarter of the total star formation budget  at $1<z<4$ is accounted for by mergers with mass ratio $>1:10$.

\begin{figure*}
\centering
\includegraphics[width=\columnwidth,angle=0]{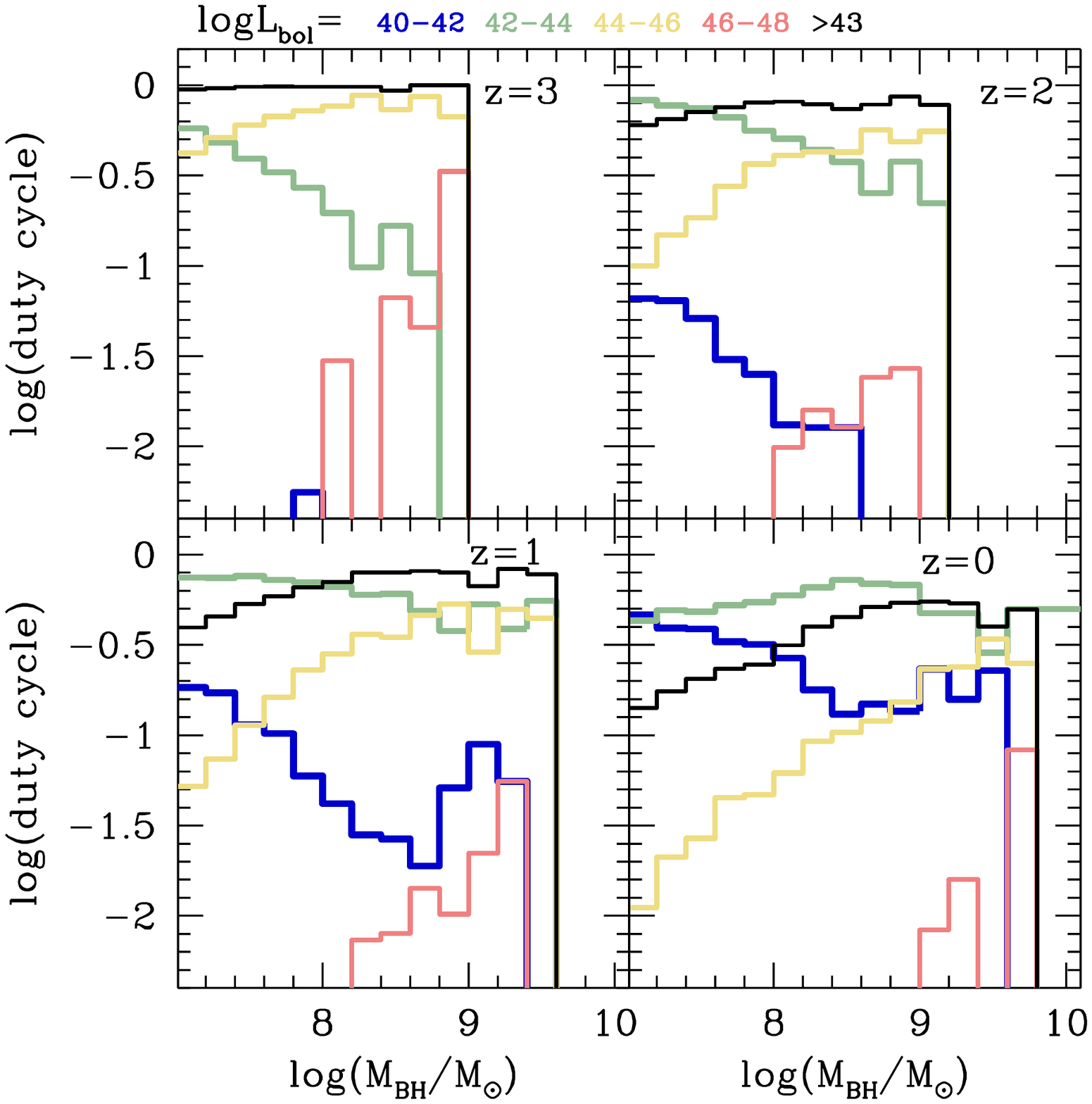}
\includegraphics[width=\columnwidth,angle=0]{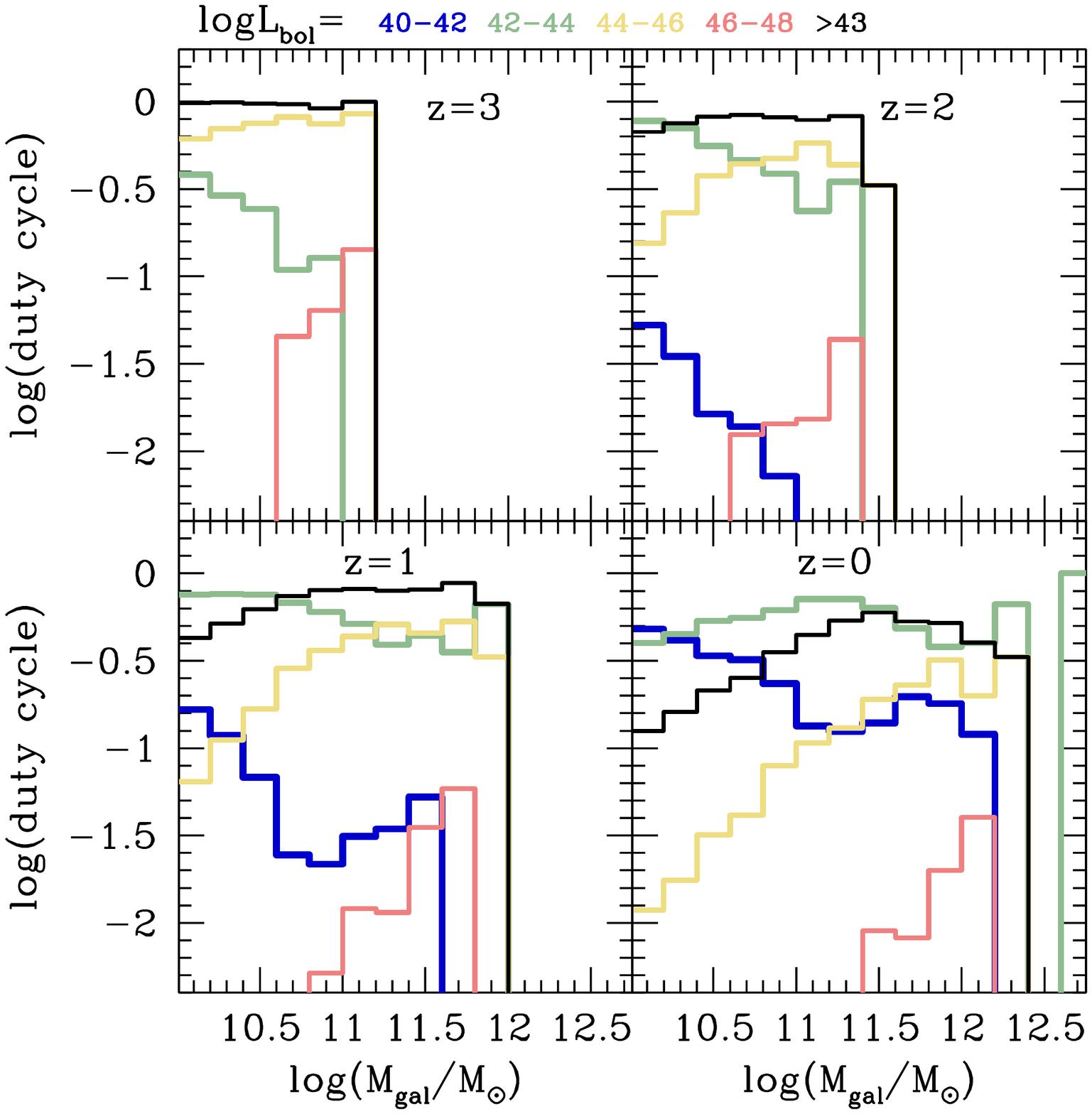}\hfill
\includegraphics[width=\columnwidth,angle=0]{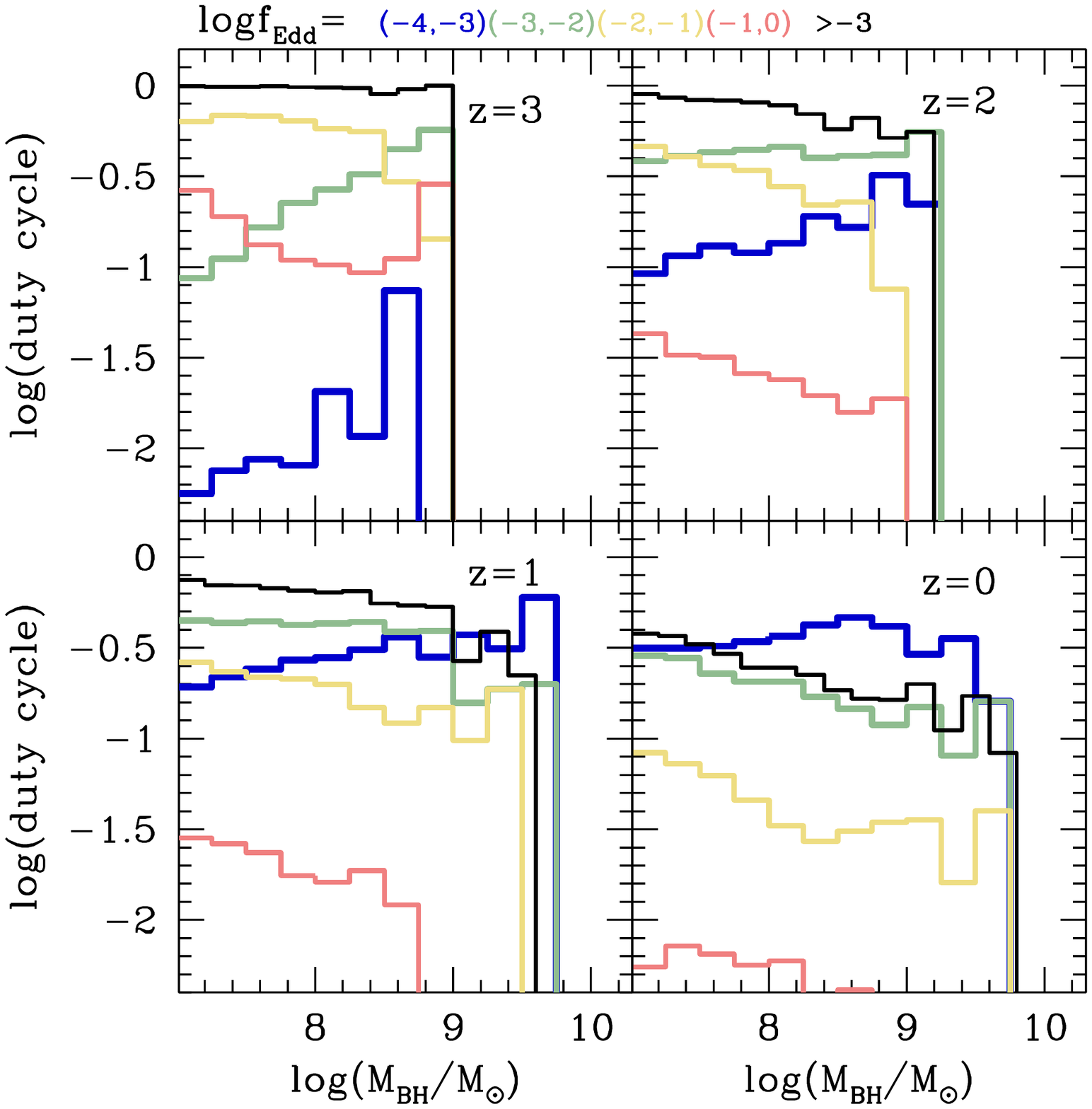}
\includegraphics[width=\columnwidth,angle=0]{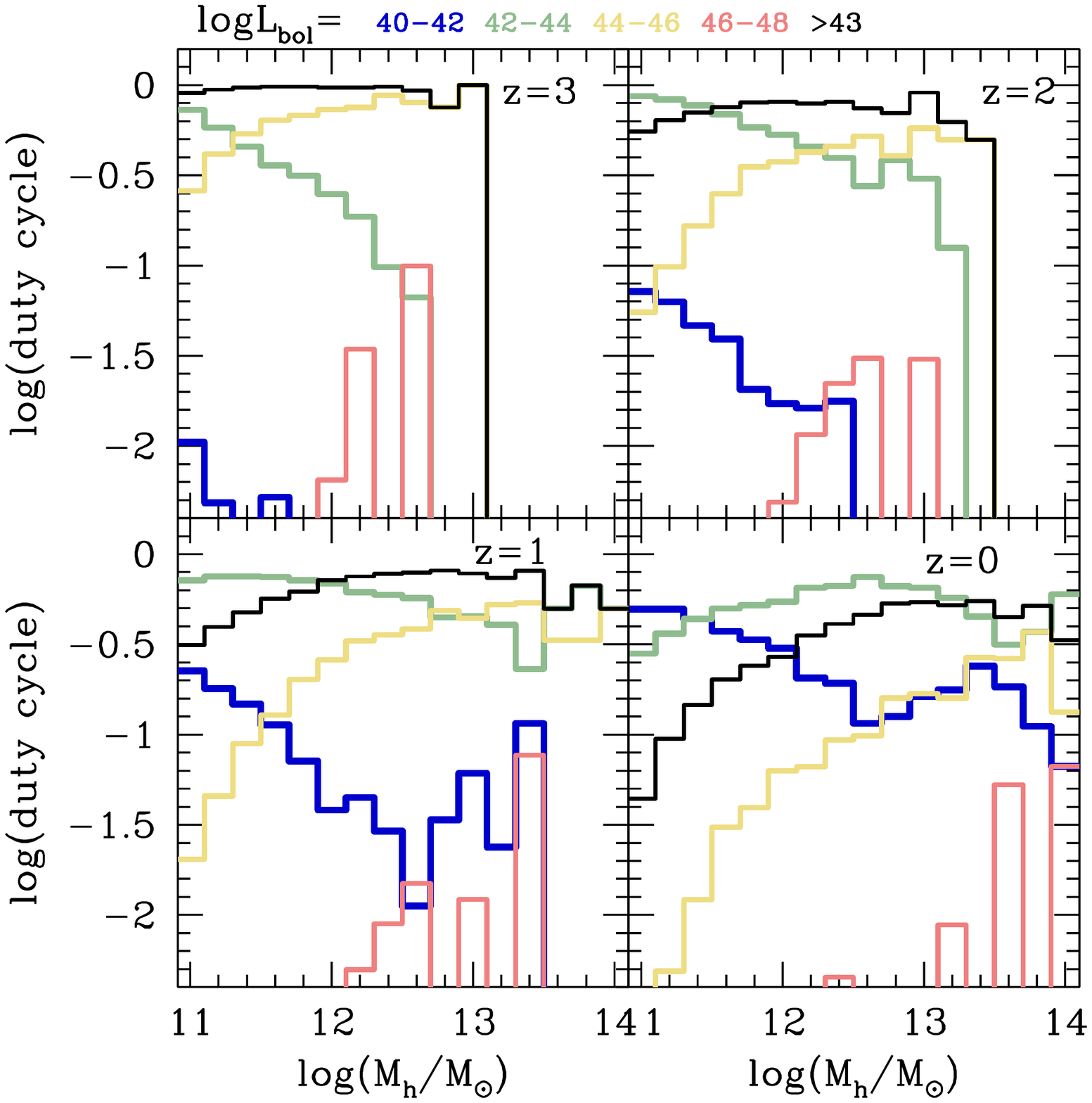}
\caption{Duty cycle, defined as the fraction of BHs in a given bolometric luminosity bin  or at a given Eddington ratio bin, over the total number of BHs (active and  inactive) in each BH/galaxy/halo mass bin.}
\label{fig:DC}
\end{figure*}

Finally, we examine the duty cycle of BHs, defined as the fraction of BHs in a given bolometric luminosity bin  or at a given Eddington ratio bin, over the total number of BHs (active and  inactive) in each BH/galaxy/halo mass bin. This is an indication of how many BHs are ``active" at a given time. As discussed by \cite{Merloni08} the duty cycle will become unity when going to sufficiently low luminosities or Eddington ratios, i.e. each BH is at some level active. We calculate the duty cycle in luminosity as a function of different quantities: BH mass, galaxy mass and halo mass. We also include the duty cycle in Eddington ratio as a function of BH mass to ease comparison with previous figures and discussions.  The latter (bottom-left) is arguably the cleanest way to highlight the cosmic downsizing, i.e., that the most massive BHs were more active at earlier cosmic times. For instance, 30 per cent of BHs with mass $\sim 10^9 \msun$ have $f_{\rm Edd}>0.1$ at $z=3$, but this fraction drops below 0.1 per cent by $z=0$.

The drop in activity, however, straddles all masses. An interesting feature is the change in gradient, at different redshifts, for some $f_{\rm Edd}$ and $L_{\rm bol}$ bins, e.g., $\log (f_{\rm Edd})\in [-3,-2)$ and $\log (L_{\rm bol})\in [42,44)$. As noted in Fig.~\ref{fig:rholum} for the integral quantity, there is a shift between fast accretors and slow accretors, and we can now track how this transition occurs at different BH masses. Unfortunately BH mass and Eddington ratio are difficult to obtain observationally, therefore we complement our analysis with additional definitions of the duty cycle, where the luminosity/galaxy mass one is perhaps the one of most immediate application for observers. Some of the trends remain in place, but information is lost for BHs in the most massive galaxies -- given their large BH mass, they appear rather luminous despite their intrinsic accretion rate being low. 

The black curve in the top-right panel can also be read as the global AGN fraction (cf. the orange solid curve in Fig.~\ref{fig:gal_OF}). The AGN fraction increases with stellar mass, and it is between 10 and 80 per cent at $0<z<1$.  This is higher by a factor of 2-3 than in X-ray selected AGN samples \citep[see][for a review]{2015A&ARv..23....1B}, while it is more similar to that of \cite{2013ApJ...764..176J}, who adopt multi-wavelength AGN diagnostics, and of \cite{2015MNRAS.447.2085S} who start from optically selected type-1 AGN and correct for obscured sources.  An important point to keep in mind is that what we estimate is the intrinsic AGN fraction, and we have not included any correction for obscuration or for Compton Thick AGN.  The obscured fraction varies as a function of redshift and luminosity, and it is estimated to be $\sim 80$  per cent at $L_{\rm bol}\sim 10^{43} \,\rm ergs$  and between $\sim 10-80$  at $L_{\rm bol}\sim 10^{45} \,\rm ergs$, depending on redshift, according to X-ray studies \citep[e.g.,][]{Lafranca2005,2014ApJ...786..104U,2014MNRAS.437.3550M}.  The population of Compton Thick AGN is highly debated, but can be as large as that of obscured AGN \citep{2015ApJ...802...89B}. Taking this into account, the AGN fraction becomes in line with observed values in X-ray selected AGN samples that find values of 5 to 30 per cent.

\section{Conclusions}\label{sec:Conclusions}

We have analyzed the evolution of BHs in a large-scale cosmological hydrodynamical simulation, Horizon-AGN.  Even in a state-of-the-art simulation like this \citep[see also][]{2012MNRAS.420.2662D,2012ApJ...745L..29D,2015MNRAS.452..575S,2015MNRAS.446..521S} internal processes in galaxies are unresolved, at the level where, e.g., star formation occurs in molecular clouds of parsec size. However, they capture well their environment, i.e. the time evolution of their inflow boundary conditions. In a sense the same applies to BHs, with the caveat that the inflow boundary conditions we use for them are less well captured: they are estimated on scales too large compared to the radius where BHs dominate the dynamics of gas (the Bondi radius)  and they depend also on the unresolved, poorly understood galaxy internal processes. Despite these shortcomings, a self-consistent AGN simulation has the advantage of better estimating how much gas gets accreted by the BH and when the correct amount of energy is dumped into the interstellar, intergalactic and circumgalactic media and, consequently, how accretion of gas on the galaxy from the cosmic web and star formation in the galaxy are affected. The back-reaction of the ambient medium on BH growth is also a crucial aspect inherently addressed.

We have first validated the simulation against a series of observables (luminosity function of AGN,  BH mass function,  BH mass density versus redshift, and relations between BH and galaxy properties), which are well reproduced for BHs with mass $\gtrsim 10^7 \msun$ hosted in galaxies with mass  $\gtrsim 10^{10} \msun$. We have then analyzed the simulation to estimate some properties of the BH population in galaxies, with an attention to quantities that can be tested in observations, and therefore provide a testbed for how simulations grow BHs in a cosmological context. We summarise our results as follows.  

\begin{itemize}
\item  The population of BHs and AGN in Horizon-AGN reproduce a variety of observational constraints, both at $z=0$ and at higher redshift, but for some notable exceptions, that we discuss below.

\item  We find a steep faint end of the high redshift AGN LF, Êbut  the  observational constraints are poor. Additionally, stronger SN feedback would suppress BH growth in low-mass galaxies and improve the overall match \citep{2015MNRAS.452.1502D}.

\item At the other extreme, for the most massive BHs, on the one hand, the bright end of the  LF (Fig.~\ref{fig:LF}) is also overestimated at $z=0.5-2$, but the errorbars are large. On the other hand, the mass function at $z=0$ is reasonable, and when we look at the BH-galaxy correlations  (Figs. ~\ref{fig:BH_gal}, \ref{fig:BH_bulge} and~\ref{fig:BH_sigma}) the simulated BHs do not seem to be more massive than the observed ones.  

\item Matching theoretical and observational samples is hard. For instance, estimating a bulge mass depends very much on the technique used, and this is true for both theory and simulations. Observed luminosity functions and mass functions are often incomplete, and derived from extrapolations of uncertain relations, therefore we have to be cautious in the comparisons.

\item In Horizon-AGN, BH masses tend to correlate tightly with total galaxy mass, and have very little scatter around that relation. This is true for simulations with similar characteristics as Horizon-AGN run with different codes. We argue that the scatter given by different merger histories is insufficient to explain the observational scatter, and unresolved internal process represent an important contribution to the scatter in the relationships  and to how well bulge properties are reproduced at fixed galaxy mass.

\item  We find that, starting at $z\sim 2$, a population of stripped sub-halos emerges. The BHs are over-massive for these halos, as their mass was set previously to the halo mass loss. Stripping, however, only seldom affects the stellar distribution. Over-massive BHs (at fixed galaxy mass) are rare in the general population, but more common in dense galaxy environments. 

\item BHs are not ubiquitous: lower mass halos and galaxies are increasingly void of BHs. This is something that has been suggested previously only through (semi)analytical models, because many simulations put BHs in all halos above a fixed mass. While Horizon-AGN does not have a very refined criterion for seeding BHs, the effect of linking BH formation to the properties of gas in galaxies, and avoiding a generalized seeding, becomes apparent.

\item Some BHs are not alone: there is a multiplicity of BHs in the most massive halos. Some of them are not associated to any subhalo and are true ``off-centre BHs". Most massive galaxies by $z=0$ host at least one off-centre BH. We identify also a population of dual AGN, where both a central and an off-centre AGN shine concurrently. This population dwindles with decreasing redshift, as also found in observations. 

\item We find very little redshift evolution in the BH-galaxy mass scaling, and a very tight relation at most cosmic times. This may be caused by the very strong self-regulation via AGN and SN feedback adopted in Horizon-AGN. 

\item We highlight transitions between fast accretors ($f_{\rm Edd}>0.1$), slow accretors ($0.01<f_{\rm Edd}<0.1$) and very slow accretors ($f_{\rm Edd}<0.01$) occurring at $z=3$ and $z=2$ respectively. 

\item We find that the Eddington ratio distribution in Horizon-AGN falls in between two different observational determinations. We find a turnover, which Aird et al. did not identify, but at lower $f_{\rm Edd}$ than found by \cite{2015MNRAS.447.2085S}. 
\end{itemize}

This paper is a general introduction to the cosmic evolution of massive BHs in Horizon-AGN. In future papers we will address more directly the growth paths of BHs and galaxies, through mergers and gas accretion, and the relation between BHs and the cosmic web.

\section*{Acknowledgements}
We thank the referee for helpful suggestions and insightful comments. We thank Val\'erie de Lapparent, Amy Reines and Francesco Shankar  for fruitful discussions, and Andreas Buchner for providing the LF complete data.  This work was granted access to the HPC resources of CINES under the allocations {2013047012, 2014047012 and 2015047012} made by GENCI. Part of the analysis of the simulation was carried out using the DiRAC facility, jointly funded by BIS and STFC. This work is partially supported by the Spin(e) grant ANR-13- BS05-0005 of the French Agence Nationale de la Recherche,  and by the National Science Foundation under grant no. NSF PHY11-25915, through the Kavli Institute for Theoretical Physics and its program `A Universe of Black Holes'. MV thanks   G.~Mamon for help with supermongo and acknowledges funding from the European Research Council under the European Community's Seventh Framework Programme (FP7/2007-2013 Grant Agreement no.\ 614199, project ``BLACK''). The research of JD is supported by Adrian Beecroft, The Oxford Martin School and STFC.
We thank  S. Rouberol for running  smoothly the {\tt Horizon} cluster for us.  

\scalefont{0.94}
\setlength{\bibhang}{1.6em} 
\setlength\labelwidth{0.0em}
\bibliographystyle{mn2e}
\bibliography{biblio}

\normalsize

\end{document}